**RESEARCH ARTICLE** OPEN ACCESS

# Backward Growth Accounting: An Economic Tool for Strategic Planning of Business Growth

Ali Zeytoon-Nejad

Wake Forest University, Winston-Salem, NC, USA

**Correspondence:** Ali Zeytoon-Nejad (zeytoosa@wfu.edu)

**Received:** 26 September 2024 | **Revised:** 17 February 2025 | **Accepted:** 24 February 2025

**Funding:** This paper received financial support from the School of Business at Wake Forest University.

## ABSTRACT

Business growth is a goal of great importance for its both private and social benefits. Many firms view business growth as an imperative for their survival, stability, and long-term success. Business growth can be socially beneficial, too, as it enables businesses to expand into new territories where they can stimulate economic growth and development, creates more jobs, increase living standards, and better serve their communities by giving back more through Corporate Social Responsibility (CSR) initiatives. Business growth must be planned reasonably and optimally so that it can effectively achieve its critical ambitions in business practice. The current common practices for planning the supply side of business growth are usually ad-hoc and lack well-established mathematical and economic foundations. The present paper argues that business growth planning can be pursued more structurally, reliably, and meaningfully within the framework of Growth Accounting (GA), which was first introduced by Economics Nobel Laureate Robert Solow to study economic growth. It is shown that, although GA was initially put forth as a procedure to explain "economic growth" *ex-post*, it can similarly be used to plan "business growth" *ex-ante* when a general backward approach is taken in its procedure—called Backward Growth Accounting (BGA) in this paper. Taking this well-established economic–mathematical approach to planning business growth will enhance the current practices conceptually and structurally, as it is built on the basis of economic logic and mathematical tools. BGA can help businesses identify and plan for key drivers of output growth and assess shortcomings in the growth process, such as poor productivity, inadequate labor utilization, or insufficient capital investment. The paper outlines an eight-step procedure for planning business growth using BGA and includes appendices with real-world examples.C5, D2, L1, M11, M21, O4.

## 1 | Introduction

Business growth can be defined as an overall increase in a company's operational scale, encompassing both increased supply and increased demand for its product. That is to say, business growth involves the growth of two distinct dimensions of business operation, namely the demand side (encompassing factors such as the customer base, quantity demanded, pricing, and revenue) and the supply side (encompassing factors such as production scale, production technology, supply capacity, and production costs).

Business growth is important simply because it is closely aligned with the firm's bottom line, which, realistically, is to maximize profit. Business growth benefits the firm through two channels of increased revenue (primarily due to a larger market share and sales) as well as decreased costs (primarily due to economies of scale and specialization). Besides the rise in profits for private firms, business growth can bring about extensive benefits to society and the macroeconomy as a whole, too. As such, business growth is a goal of great importance both privately and socially, and it must be planned reasonably and optimally to succeed.

However, the current common practices for planning the supply side of business growth are usually ad-hoc, baseless, and lack well-established, mathematical foundations. The present paper argues that business growth planning can be pursued more structurally, reliably, and meaningfully within the framework of Growth Accounting (GA) – as introduced by Robert Solow (1956, 1957), a winner of the Nobel Prize for Economics in 1987 – when a backward general approach is applied to it.[1] The present paper reverse-engineers GA and makes an application of it called Backward Growth Accounting (BGA) that can be effectively used in business practice to navigate the process of business growth systematically. BGA provides an instrumental framework to plan the supply side of business growth.

GA is an economic methodology to explain the factors that contribute to economic growth. In other words, GA is a mathematical method to decompose growth in output into several sources of growth including "growth in the stock of inputs"[2] and "growth in the productivity of inputs". However, this paper proposes a backward approach to GA which can be applied in business practice to plan business growth optimally and effectively. The usefulness of this backward approach to growth accounting is discussed, illustrated, and exemplified throughout the paper. This study addresses five key questions:

- What are the critical factors that need to be considered when planning growth for the supply side of a business?
- How can inputs be structured in an input–output setting to facilitate the mathematical planning and navigation of business growth?
- How can the GA methodology, originally introduced in economics, be modified in its general approach to meet the needs of businesses and aid in their business growth planning efforts?
- What general strategies with respect to sources of business growth can be considered when planning for growth on the supply side of a business?
- How can BGA be applied in business practice to provide concrete, actionable insights and propose practical, strategic moves for business planning purposes?

This paper answers the above questions and introduces BGA as a new general, mathematical approach with respect to planning business growth, which will be applicable to any types of business producing any types of goods or services. Business leaders can utilize BGA as a vital tool to pinpoint the key drivers of their output growth, enabling them to optimize their utilization and effectively plan for business expansion. This approach aids in identifying how changes in input quantity and productivity impact output growth, allowing for the determination of significant components contributing to overall output growth rates. Through BGA, business analysts can assess the shortcomings in the output growth process, such as inadequate labor utilization, insufficient capital investment, or poor productivity, and identify the most viable solutions to address these issues.

The remainder of the paper is organized as follows. The next section provides a formal exposition of GA first. Afterwards, it explains the methodology of GA in detail and provides examples to illustrate how GA is used to study economic growth. Sections 3 and 4 discuss the importance of business growth and planning business growth, respectively. In Section 5, BGA is introduced and formulated as an instrumental framework for planning business growth. In Section 7, a conclusion will be drawn from the whole discussion and the main points and considerations are summarized. Finally, the paper will end with appendices to illustrate and exemplify the BGA method in greater detail in order to show some real-world applications of this approach more clearly.

## 2 | Growth Accounting: A Well-Established Tool to Study Output Growth

Growth accounting is a procedure used in economics to measure the contributions of different factors to economic growth taking place over time in an economy. Growth accounting is traditionally used as a tool to identify the sources of growth in aggregate output in the economy. In that context, economic growth is measured as percentage change in real output. Growth accounting decomposes this rate of growth to its different sources, including "growth in the stock of inputs" and "growth in the productivity of inputs". By doing so, one can attribute separate portions of output growth to different contributing factors to growth.

One of the main advantages of growth accounting is to help in teasing out the effect of the improved productivity of inputs on output growth. This task is particularly important because the improvement in total factor productivity is usually unobservable in real-world phenomena, while it typically accounts for a significant portion of output growth in developed and developing economies in the twentieth century (e.g., see Solow 1957; Abramovitz 1993; Crafts 2003; Field 2006; Crafts and O'Rourke 2014; Crafts and Woltjer 2021; Pan et al. 2022). In what follows, the mathematical construct of growth accounting is shown and the way it estimates the share of improved productivity of inputs in output growth is illustrated in a simple example.

The first step in growth accounting is to identify the inputs and output involved in the production process of interest. Here, to simplify matters, a process of production with two classical inputs in economics, namely labor and capital, are considered without loss of generality. However, one can include as many inputs as they find relevant in their respective process of production. Next, the production function underlying the process of production under study must be specified and estimated. A production function shows the mathematical relationship between the output quantity of a process of production and the quantities of the inputs used to produce that output. The generic form of a production function is as follow:

$$Y = TFP \cdot f(L, K) \quad (1)$$

where $Y$ denotes output quantity, $TFP$ denotes the level of total factor productivity[3] at time $t$, and $L$ and $K$ denote the stock of Labor and Capital, respectively, engaged in the process of production. This generic form of production function can take different specific forms depending on several considerations such as the relationship between the inputs of production (i.e., whether they are perfect complements, perfect substitutes, or relative substitutes). A common practice in growth accounting is to assume the Cobb–Douglas (CD) production function as the specification of production primarily because of its many desired mathematical properties, theoretical support, and estimation flexibility.[4] Following this common practice in growth accounting, the CD production function can be written as follows:

$$Y = TFP \cdot L^{\alpha} \cdot K^{\beta} \quad (2)$$

where $\alpha$ and $\beta$ are the parameters associated with the stock of labor and capital, respectively. It is proven for the case of CD production function that $\alpha$ and $\beta$ are also equal to "labor elasticity of output" and "capital elasticity of output", respectively. Now, one can take the natural logarithm of both sides of this equation and end up with the following equation:

$$\ln Y = \ln TFP + \ln L^{\alpha} + \ln K^{\beta} \quad (3)$$

Using the well-known mathematical property of the natural logarithm function (which says $\ln X^{\alpha} = \alpha \cdot \ln X$), the above equation can be re-written as the following equation:

$$\ln Y = \ln TFP + \alpha \cdot \ln L + \beta \cdot \ln K \quad (4)$$

Now, one can take the derivative of both sides of the equation with respect to time $t$, and write the resulting equation in a discrete form as follows:

$$\frac{\Delta Y}{Y} = \frac{\Delta TFP}{TFP} + \alpha \cdot \frac{\Delta L}{L} + \beta \cdot \frac{\Delta K}{K} \quad (5)$$

In a more concise form, this equation can be written as the following equation, which is known as the fundamental equation of growth accounting:

$$g_Y = g_{TFP} + \alpha \cdot g_L + \beta \cdot g_K \quad (6)$$

where $g_Y$ refers to growth in output, $g_L$ denotes growth in labor (i.e., growth in the number of workers), and $g_K$ denotes growth in capital (i.e., growth in the stock of capital, which can include growth in the stock of machinery, equipment, buildings, etc.), and $g_{TFP}$ denotes Total Factor Productivity Growth (aka TFPG), which technically shows the portion of output growth that cannot be attributed to the accumulation of factors of production (i.e., labor and capital growth), so it must be attributed to the improved productivity of factors of production, which is called TFPG.[5]

In the fundamental equation of growth accounting, $\alpha$ and $\beta$ are identified from the estimation of production function on the basis of the real-world data available on $L$, $K$, and $Y$. Using the same dataset, $g_Y$, $g_L$, and $g_K$ can be computed, and the only unknown in this equation will be $g_{TFP}$, which is estimated as the residual of this equation as follows:

$$g_{TFP} = g_Y - \alpha \cdot g_L - \beta \cdot g_K \quad (7)$$

This is the reason why the term $g_{TFP}$ or TFPG is also known as the Solow residual, which is interpreted as the portion of output growth that is not explained by the increased amounts of inputs used in the process of production. That is, TFPG refers to the portion of output growth that cannot be attributed to the accumulation of factors of production; as such, the only remaining explanation for this portion of output growth is an increase in the productivity of production factors.[6]

According to the fundamental equation of growth accounting for the case of two classical inputs of $L$ and $K$ (i.e., equation (6)), the different sources of output growth can be listed as the following:

1. The portion of growth that is achieved due to "an increase in the number of workers" ($g_L$), in which case the share of this contributing factor to output growth is equal to "$\alpha \cdot g_L$".

2. The portion of growth that is achieved due to "an increase in the stock of capital" ($g_K$), in which case the share of this contributing factor to output growth is equal to "$\beta \cdot g_K$".

3. The portion of growth that is achieved due to "an increase in the level of (production) technology" or "an increase in the level of TFP", in which case the share of this contributing factor to output growth is equal to "$g_{TFP}$".

In fact, in the first two cases, the "quantity" of the factors of production has increased, while, in the third case, the "quality" or "productivity" of the factors of production has increased.

Figure 1 depicts the shares of these contributing factors to output growth in a schematic diagram.

In Figure 1, suppose that the circle on the left-hand side represents the size of output ($Y_1$) in Year 1. Furthermore, assume that the size of output in this hypothetical economy has grown over time from Year 1 to Year 2 to the size of the larger circle ($Y_2$) on the right-hand side (which is covered from inside by the original circle ($Y_1$) from Year 1). The narrow-band part of the larger circle in Year 2 in fact represents the growth in output from Year 1 to Year 2, which is equal to ($Y_2 - Y_1$) in absolute terms. Considering this change in output proportionate to the original output, the rate of change in output can be thought of as $g_Y$, which in turn can be illustrated as the narrow-band part of the larger circle in relative terms with respect to the original output. As shown in the lower part of the figure, this narrow-band part of the larger circle can be decomposed into three separate sections representing the shares of each contributing factor to output growth, including "$\alpha \cdot g_L$", "$\beta \cdot g_K$", and "$g_{TFP}$". (In absolute terms the shares of these three factors will be equal to $\alpha \cdot \frac{\Delta L}{L} \cdot Y$, and $\beta \cdot \frac{\Delta K}{K} \cdot Y$, and $\frac{\Delta TFP}{TFP} \cdot Y$, respectively.) To provide clarity and illustrate how these shares are computed, a numerical example for growth accounting is presented in Appendix 1. In addition, Figure 2 depicts the workflow of different steps in the GA procedure in a sequential manner.

An additional common practice in growth accounting is to assume the common-sense property of Constant Returns To Scale (CRTS) for production function. This property is imposed as a mathematical constraint in the econometric estimation process of production function usually as an assumption of conventional wisdom. The CRTS property requires that the production function exhibits the typical real-world situation where doubling the usage of all inputs (e.g., $L$ and $K$ in the above derivations) will also double output ($Y$). This property holds when the weights of relative importance or input elasticities of output (i.e., the parameters in the exponents) add up to unity. That is, $\alpha + \beta = 1$. Accordingly, when $\alpha + \beta < 1$, the production function exhibits the property of Decreasing Returns To Scale (DRTS), and when $\alpha + \beta > 1$, the production function exhibits the property of Increasing Returns To Scale (IRTS). Typically, these two last conditions are relatively rare to observe in real-world phenomena especially over the second stage of production, within which stage a rational producer would always choose to operate. This is the reason why the property of CRTS is usually assumed in the GA procedure as an assumption of conventional wisdom about the proportionate, harmonic relationship between changes in inputs and change in output.

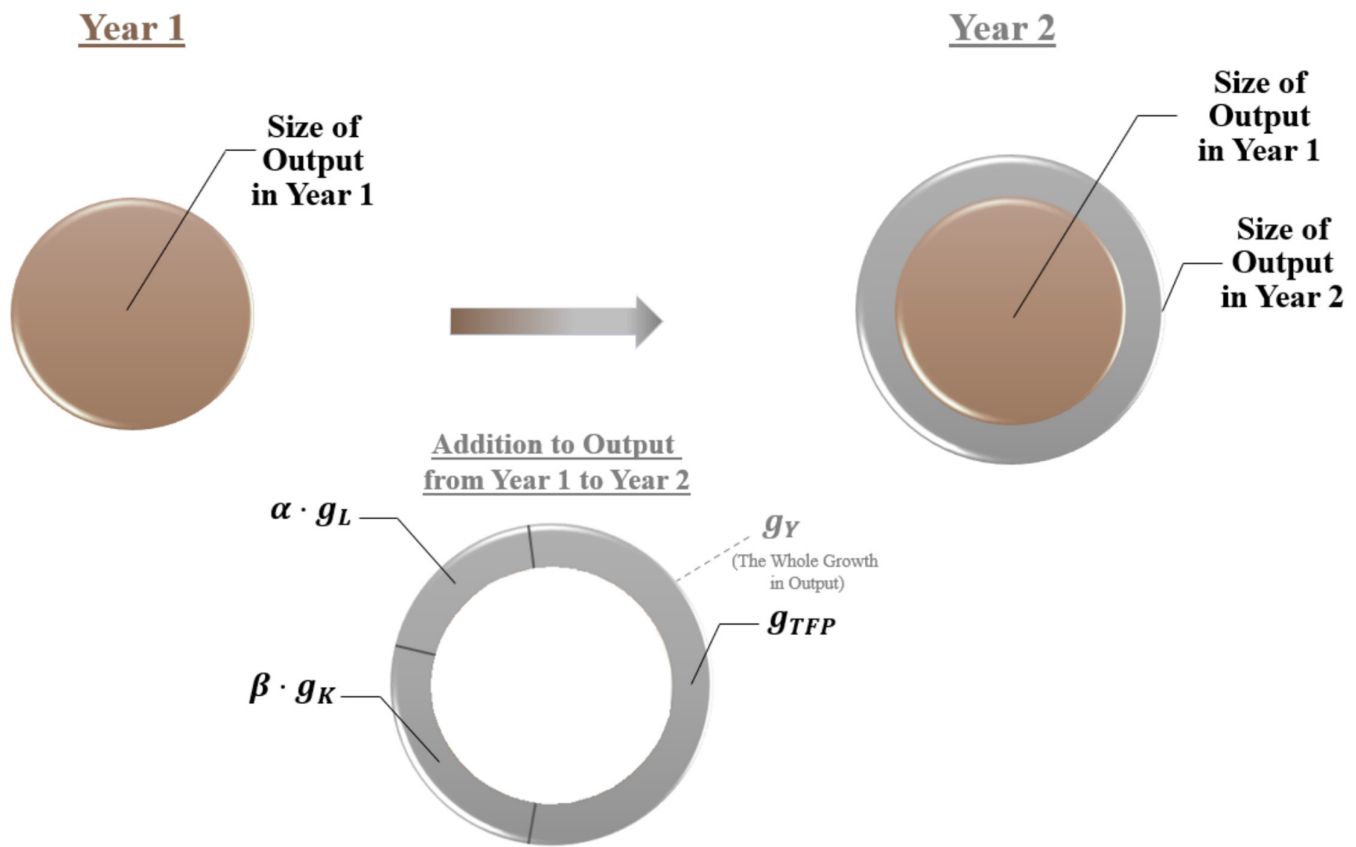

**FIGURE 1** | A schematic diagram illustrating growth in output over time and the decomposition of output growth into three different contributing factors to output growth.

Finally, a few considerations and important notes are mentioned about growth accounting in several bullet points below:

- **(On the Choice of Cobb–Douglas Production Function and CRTS)**. It is important to note that, although the Cobb–Douglas production specification and CRTS were assumed in the derivations shown and examples solved in this paper (primarily for the sake of pining down the structure choice when there was a menu of choices to choose from), these two choices are not limiting factors for the GA procedure nor for the BGA procedure (which is to be introduced in Section 5).

That is to say, although the Cobb–Douglas production function and CRTS were assumed here in accordance with common practices in GA, one can choose to apply the GA and BGA methodology with other reasonable specifications of production functions and other justifiable forms of returns to scale without loss of generality of the GA and/or BGA approach. In a sense, the default structure to start with should be Cobb–Douglas (for its desired properties and flexibility) and CRTS (for its realistic structure) unless one has a strong justification to go away from these two structures. After all, these two structures should not be viewed as limitations for the GA and BGA procedure, as one can easily choose to go with other production functions and other properties such DRTS or IRTS when using GA and BGA. These other forms are allowed by GA and BGA, and these approaches can easily accommodate such uncommon forms, but such deviations from CRTS require strong rationale and valid justifications as to why they have been chosen over CRTS. For example, one may make the case that their process of production is in the first or third phase of production where one is likely to observe IRTS and DRTS, respectively.

- **(On the Number of Inputs Included)**. Although in the above exposition and derivations of growth accounting equations, only two (traditional) factors of production were included for the sake of simplicity and brevity, the number of inputs included in the GA procedure can easily be increased to include additional types of inputs in order to be more realistic. For instance, L can be decomposed into different types of labor (e.g., $L_1$ being office staff, $L_2$ being general labor, and $L_3$ being professional employees) and K can be decomposed into different types of capital (*e.g.*, $K_1$ being the number of lines of production, $K_2$ being factory buildings, and $K_3$ being vehicles). Moreover, additional forms of factors of production can also be included depending on the nature of production under study. Examples include raw materials, parts, land, utilities, and chemicals.

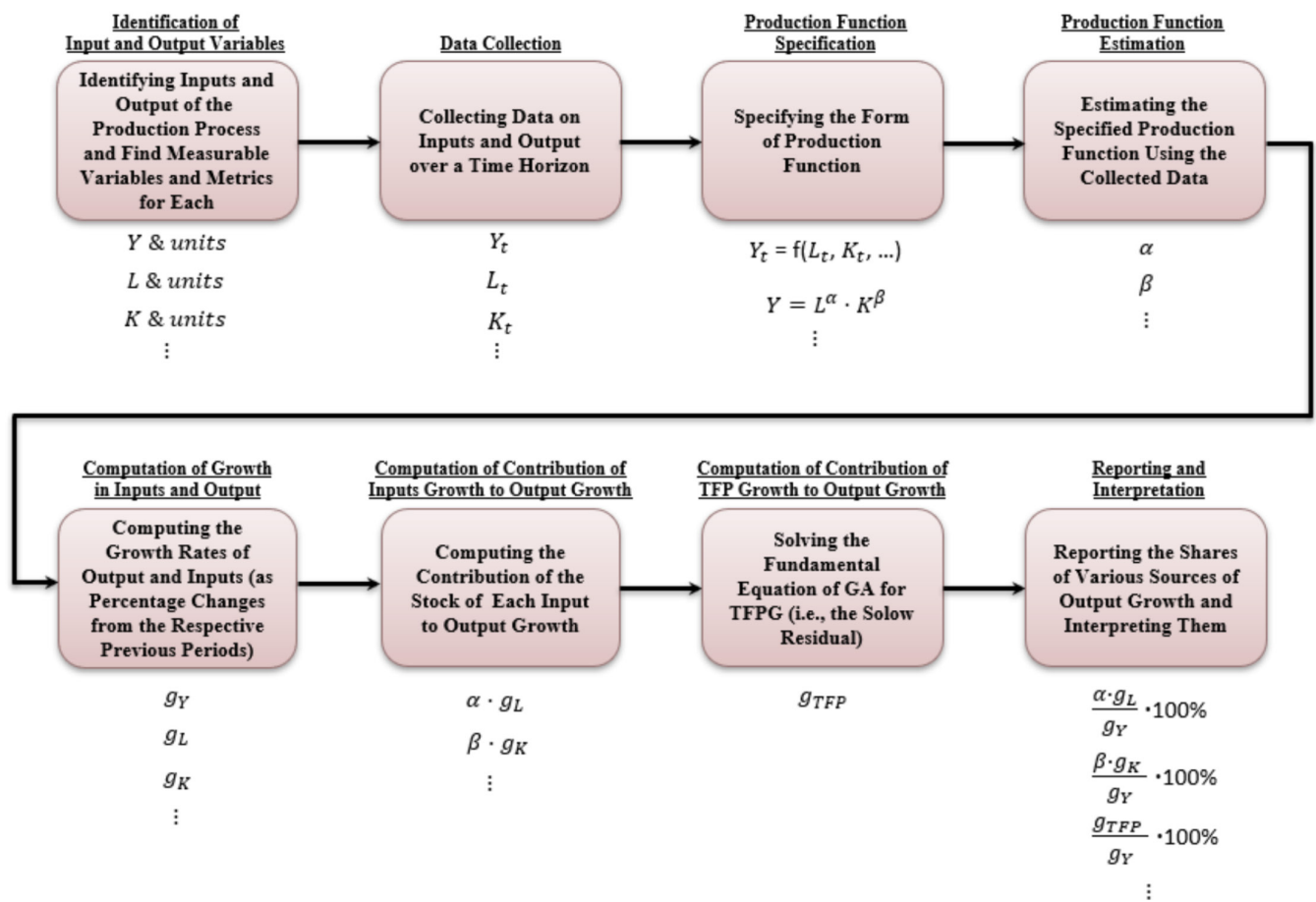

**FIGURE 2** | Workflow of sequential steps in the growth accounting (GA) procedure. *Source:* Author's Own Drawing.

- **(On Aggregate vs. Disaggregate Output Growth)**. Although GA has traditionally been used for studying output growth at an aggregate level mostly, it can also be used to study output growth at a disaggregated level such as the firm level. The GA approach conducted at a disaggregated level such as that of firm will result in even more reasonable, accurate, and reliable findings because the GA approach at the firm level to study business growth does not suffer from the aggregation problem. This advantage applies also to the BGA approach, which is to be introduced and proposed in this study in the following sections.

- **(On the Ex-Post vs. Ex-Ante Analysis of Output Growth)**. The original GA methodology provides an *ex-post*, positive (as-is) economic analysis *describing* how an economy has grown. As such, GA is a descriptive analysis that is conducted after the fact to explain how the world works (in particular, it explains the factors that have contributed to economic growth). Alternatively, and somewhat similarly but conversely, the GA approach can be modified and reverse-engineered (which will be called the BGA approach in this paper) to provide an *ex-ante*, normative (to-be) economic analysis *prescribing* how a business leader should plan to grow the supply side of their business. As such, BGA is a prescriptive analysis that is carried out before the fact to prescribe how the world should work (in particular, it prescribes to what extend and in what proportions inputs need to growth at each period of time in the future in order to ensure that the firm achieves its target growth rate of business while maintaining efficiency with respect to the relative usage of inputs).

In the next two sections, the importance of business growth and planning for business growth are discussed. In the section following those, the BGA approach, which is basically an ex-ante, normative economic analysis prescribing how the firm at a disaggregated level should plan to grow the supply side of its operation.

## 3 | Business Growth: A Path to Survival and Stability in Competitive Markets

Business growth can be defined as the overall increase in a company's operational scale, encompassing both increased supply

5 of 22

and increased demand for its product. That is to say, business growth involves the growth of two distinct dimensions of business operation, namely the demand side (encompassing factors such as the customer base, quantity demanded, pricing, and revenue) and the supply side (encompassing factors such as production scale, production technology, supply capacity, and production costs). Business growth is important because it is closely aligned with the firm's bottom line, which, realistically, is to maximize profits. Business growth benefits the firm primarily through increased revenue (due to a larger market share and sales) as well as decreased costs (due to economies of scale and specialization). Besides the rise in profits for private firms, business growth can also bring about extensive benefits to society and the macroeconomy as a whole.

Business growth can be beneficial to private firms in different ways. Examples of such benefits include: Increased financial stability (by enabling businesses to better diversify their investment portfolios and thereby spread and reduce the risk of their operations); increased likelihood of the survival of businesses (by providing them with a larger safety cushion protecting them more effectively against bankruptcy in face of exogenous shocks that may occur in the economy); strengthened competitive advantage (by enabling firms to catch up with their rivals who are investing in growth and helping the firms introduce new products or services, and enter new markets); enlarged market share; improved brand recognition and enhanced market penetration; increased cost-savings due to economies of scale (by allowing firms to negotiate better prices with their input suppliers when buying their inputs in bulk, reduce production costs through work division and specialization, have better access to financing and cheaper capital, and reduced costs of logistics, insurance, and promotion); greater opportunities for innovation and creativity (as companies invest in new technologies, research and development, and other such initiatives); greater opportunities to invest in further expansion strategies by the firm itself; attracting more investors to invest in the firm's growth plans; attracting top talents with high-quality human capital to work for the firm as employees; and finally, increased valuation of the firm.

Business growth can be socially beneficial, too, as it enables businesses to expand into new territories where they can stimulate economic development and growth; provide additional resources to improve their customer-service efforts; increase the number of customers and communities served; increase the range of products and services offered to the society's members; to better serve their communities by giving back through Corporate Social Responsibility (CSR) initiatives; and finally, increased business sustainability (by allowing firms to make long-term growth plans without any negative social, environmental, and cultural impacts and focus on the three pillars of economic viability, environmental protection, and social equity). Business growth can also have substantial positive macroeconomic effects, as it creates more jobs, stimulates economic activity, provides more trade opportunities, boosts the local and national economy, and contributes to overall economic growth and increased living standards.

In order to ensure their survival and long-term success, it is imperative for firms to view business growth as a crucial aspect of their strategy. Failing to prioritize business growth often leads to companies falling behind their competitors who are investing in their expansion efforts (e.g., see Bonaccorsi 1993; Moen 1999; Lee 2009; Çetinkaya et al. 2019; Lafuente et al. 2020). Inertia can quickly cause a decline in sales and customer retention, thereby putting the firm at risk of bankruptcy in a highly competitive market. To avoid this, companies must invest resources into achieving business growth that will position them for stability in the long run. However, achieving desirable business growth requires a realistic and effective plan. The next section delves into the significance of business growth planning.

## 4 | Planning Business Growth: Having a Plan for Growth Is Better Than Operating on a Whim

As Benjamin Franklin once put it, "If you fail to plan, you are planning to fail." This statement holds true especially in the competitive world of business, where robust planning is the cornerstone of growth and success. Planning for business growth might seem like a daunting task, but it is a necessary process that ensures economic sustainability in competitive markets and can considerably help businesses in achieving their growth ambitions. Business growth planning is vital for several reasons. It identifies aims and objectives, provides direction and focus, mitigates risks and uncertainties, facilitates informed decision-making, fosters communication and collaboration, and helps in measuring progress and track growth and success.

The first step in planning business growth is to clearly identify the objectives of growth. Then, one can set a target growth rate and accordingly plan the different dimensions of business to achieve the designated target. Businesses that have clearly defined objectives and set a growth target to achieve those objectives can develop a roadmap accordingly for their target growth. This makes them far more likely to achieve their objectives than those that operate on a whim. Planning enables businesses to prioritize their goals more realistically, allocate resources more efficiently, and focus on activities that will serve their business growth ambitions more effectively. A well-developed plan for business growth will outline the strategies that a business must implement to overcome its weaknesses and leverage its strengths in the direction of growing the business towards its designated target. Such a roadmap for business growth will help steer the firm towards growth more successfully.

In addition, planning business growth helps businesses mitigate risks and uncertainties. No matter how well-prepared a business is, risks are inevitable, especially when it comes to the inherently dynamic process of business growth. However, with a well-thought-out plan for business growth, businesses can minimize the impact of risks. A business growth plan will identify potential risks and provide strategies to address them, ultimately reducing the likelihood of failures that may damage the firm's financial stability and existence. Besides, planning business growth facilitates informed decision-making about business growth. In an ever-changing market environment, the ability to make well-informed decisions is critical. Planning business growth involves analyzing historical and real-time data and conducting research to identify trends, patterns, and emerging opportunities. This information empowers business owners to make informed decisions about business growth, which align with their goals while minimizing errors and potential losses. Therefore, a well-developed plan for business growth will enable businesses to identify issues before they occur, develop contingency measures, and make better decisions.

Furthermore, business growth planning fosters communication and collaboration. Effective planning requires collaboration and communication within the firm's overall organization. Planning business growth will also enable teams to identify and solve issues around growth together, foster innovation for growth, and improve overall growth performance. Moreover, business growth planning enables businesses to measure progress and track growth and success. A well-developed plan for business growth involves setting targets, benchmarks, and growth performance indicators that enable businesses to track their progress towards achieving their growth ambitions. With this approach, businesses can easily identify the areas in the growth process where improvements are required to achieve their objectives.

It is also important to note that an effective plan for business growth must be formulated in alignment with other existing plans that a business operates based on, including the overall business plan, strategic plans, marketing plans, and succession plans. While planning might seem intimidating, it is a necessary process that ensures survival in highly competitive markets. Therefore, businesses must prioritize planning for business growth and allocate resources to develop robust and well-defined growth plans that will enable them to achieve their growth objectives. The present paper proposes an approach to planning business growth which is derived from the GA approach. Although GA was initially put forth as a procedure to explain "economic growth" *ex-post*, it can similarly be used to plan "business growth" *ex-ante* when a general backward approach is taken in its procedure. This approach will be explained in great detail in the next section.

## 5 | Backward Growth Accounting: An Instrumental Framework to Plan Business Growth

Although planning is key to business growth and success, current practices to plan for business growth are usually theoretically ad hoc. To improve these current practices structurally and theoretically, it will be very useful and greatly advantageous to take a well-established, economic-mathematical approach to planning business growth. In this section, it is argued that GA (which is an extensively discussed practice in the literature of the theory of economic growth) can provide such a useful framework for planning business growth. However, the sequential procedure of the GA method must be slightly tweaked and re-ordered so as to account for the *ex-ante* nature of planning business growth. In particular, the GA approach needs to be pursued "backward" in order for it to become *ex-ante* and applicable to the nature and purpose of planning business growth. As such, this backward approach to the GA is called Backward Growth Accounting (for short, BGA) in this paper.

Figure 3 depicts the workflow of different steps in the BGA procedure in a sequential manner below.

As shown in Figure 3, conducting a BGA analysis requires taking a series of steps. It is important to note that the first row of this visual workflow (which illustrates the phase of identification, specification, and estimation of the production function) is the same as the first row of the workflow of the AG procedure. However, the steps listed in the second row (which illustrates the phase of growth-accounting computations and interpretation) follow a procedure that entails a backward approach to the general procedure of the ordinary GA. All these steps are explained one by one in what follows.

- **Step 1:** The first step of the BGA is to identify the inputs and output of the production process under study. Then, measurable proxy variables and physical dimensions or units are needed to be defined for each input and output.[7]
- **Step 2:** A dataset needs to be collected including all the input variables and output variable involved in the production process of interest. The data will be collected for a sufficiently large dataset (temporally, or cross-sectionally, or in a panel form), so that it can be used to estimate the production function reliably.
- **Step 3:** An appropriate form of a production function will be specified, which not only considers the type of the relationship between inputs (i.e., whether they are substitutes, complements, or relative substitutes), but also has desired mathematical properties, theoretical support, and estimation flexibility, as discussed in great detail in Section 2. As mentioned earlier, a common practice in GA is to go with the Cobb–Douglas production specification for the reasons mentioned in Section 2. The same practice can be followed in the BGA procedure with the same rationale.
- **Step 4:** Using appropriate econometric methods and the dataset collected, the specified production function will be estimated, and the power parameters of the function will be estimated.
- **Step 5:** In this step, the objectives of the growth plan are clearly defined, and a realistic target timeframe to achieve the objectives are set. Then, the required rate of growth in output to achieve the set objective over the set timeframe is computed.
- **Step 6:** In this step, a realistic strategy plan for the designated growth rate is formulated, which needs to outline from which growth sources the set output growth is planned to be achieved. In general, there are three possible ways to achieve output growth according to the literature of GA in economics, which are as follows:
  ◦ Increasing the quantity of the factors of production (i.e., increasing the quantity of labor and capital), which is also called input accumulation.
  ◦ Improvements in the productivity of factors of production (i.e., increasing the productivity of labor and capital – while keeping the quantity constant), which is also called Total Factor Productivity Growth (TFPG).
  ◦ A mixed combination of input accumulation and TFPG.

Figure 4 illustrates these three approaches to output growth in detail (in a single-input case), where the three diagrams show output quantity (*Y*) on the vertical axis vs. labor quantity (*L*) as an input on the horizontal axis *ceteris paribus*.

Working on a two-dimensional space to model the input–output relationship in this series of diagrams, consider labor (*L*) as the only input in this single-input representation. As shown in this figure, there are three general approaches that can be taken with respect to planning output growth. Suppose the set target rate of output growth here is equal to 20% in a certain period of time (say, one year), which is depicted as a 20% increase in *Y* from $Y_A$ to $Y_B$ on the vertical axis in all the three diagrams. Now, this increase in *Y* can be obtained in three different ways, which are illustrated in the three diagrams.

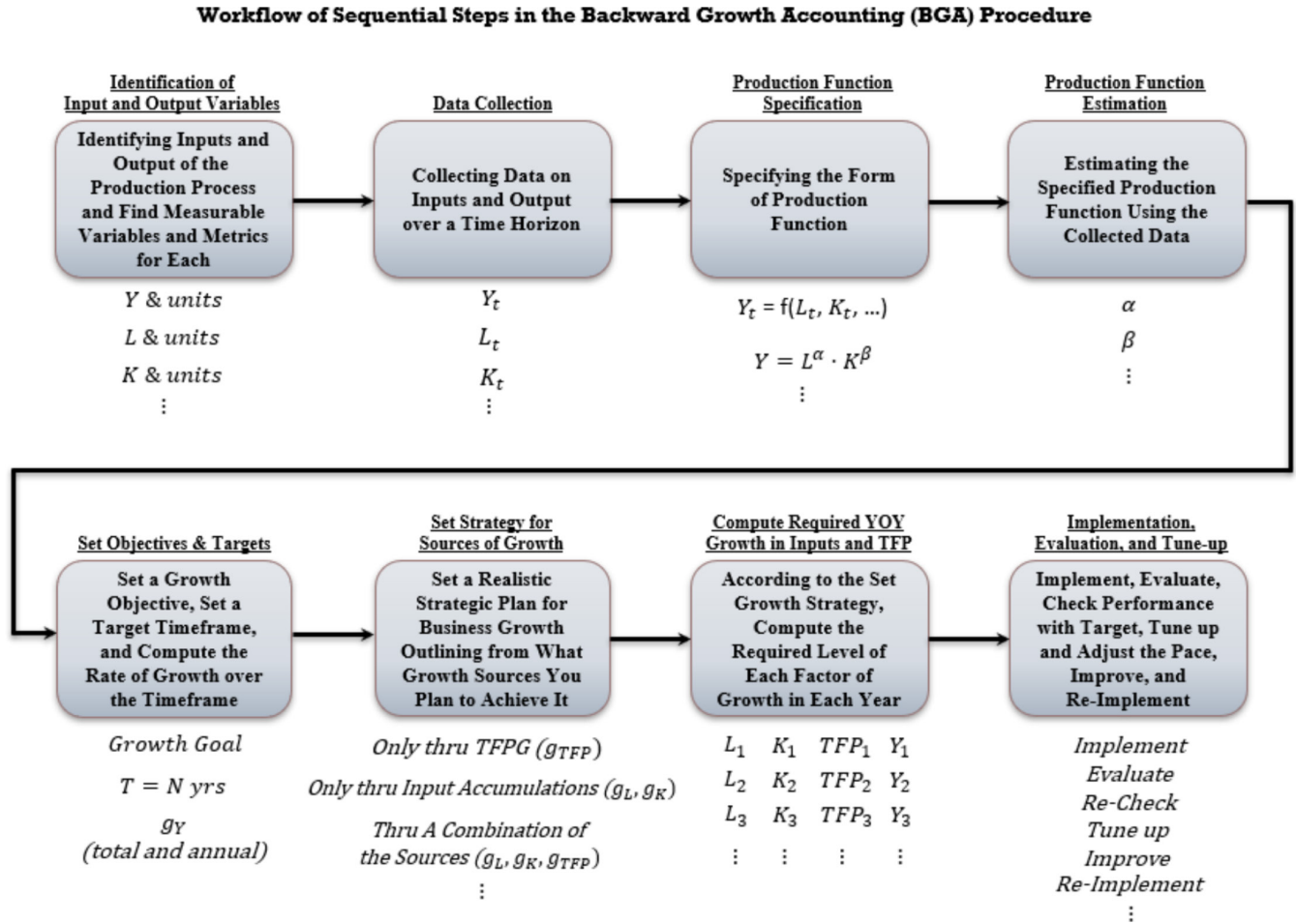

**FIGURE 3** | Workflow of sequential steps in the backward growth accounting (BGA) procedure. *Source:* Author's Own Drawing.

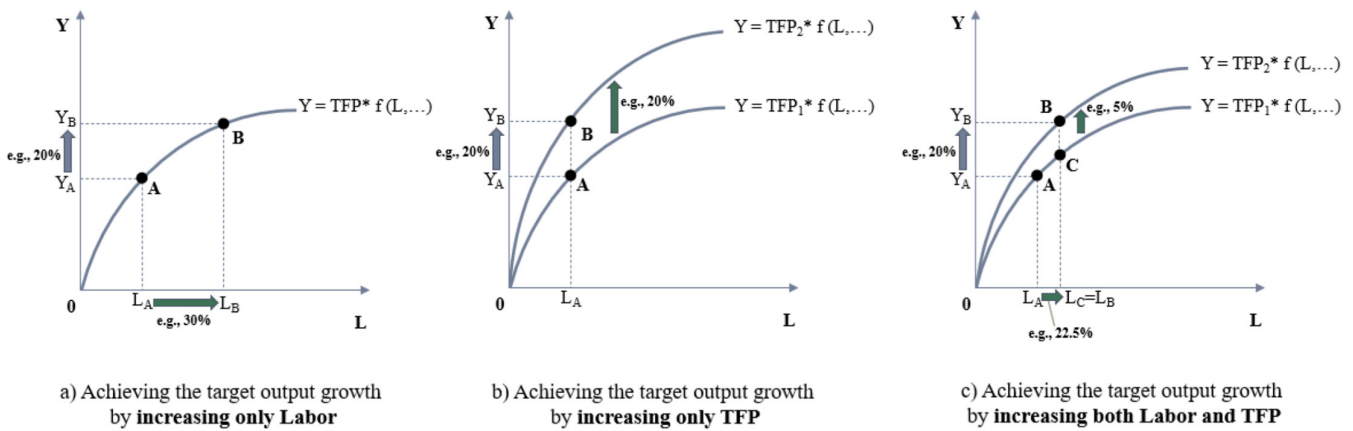

**FIGURE 4** | Three possible strategies for determining sources of growth in planning business growth. *Source:* Author's Own Drawing.

In the first diagram on the left, it is shown that one way to attain this amount of growth is to increase the quantity of labor by 30% (i.e., hiring 30% more workers). Assuming that the marginal contribution of labor growth to output growth (that is, $\alpha$) is equal to 2/3, the 30% increase in $L$ will result in 2/3*30% growth in output, which is equal to 20% growth in output. In this case, the target growth rate in output is achieved by moving along the same production function, while keeping the level of TFP unchanged.

An alternative way to achieve the 20% target rate of growth in output is to increase the TFP level by 20% alone, while keeping the same number of workers, as shown in the middle diagram in Figure 4. In this case, output growth can be accomplished by keeping the same number of workers, but reorganizing the business and workers for more efficiency, reassigning and reshuffling tasks to improve performance, enhancing the status of work division and specialization, and/or somehow empowering the workers and making them more productive, say, by having them pursue certain work-related educations, or receiving additional work training, or providing them with more incentives (e.g., by implementing a pay-for-performance compensation instead of a fixed salary). This type of growth in TFP is also known as "technical progress" in production process. In this case, the target output growth is achieved through shifting the production function up by 20%, while keeping the number of workers the same as before.

More realistically, a firm can choose to follow a mix of these two growth strategies, which is depicted in the diagram on the right-hand side of Figure 4. In this case, both channels of "input accumulation" and "TFPG" are being utilized simultaneously to achieve the 20% target rate of output growth. That means, under this strategy, part of output growth will come from input accumulation and moving up along the production function (e.g., 15% of growth will come from hiring more workers—because 2/3*22.5% = 15%) and the remaining target output growth (i.e., the remaining 5%) will come from TFPG and shifting the whole schedule of the production function up by 5%. This third strategy plan may be more desired and more realistic in many cases. This is because productivity growth is a potential source of output growth, and as such, it should be taken advantage of when it is available as a source of economic growth to the firm, especially in cases where it is more efficient (cost-wise) to achieve business growth through that channel. At the same time, the capacity of growth through solely productivity is sometimes limited, especially over short timespans. TFPG occurs most reliably over the long run. In that case, the other source of output growth must be utilized, which is a proportionate accumulation of all inputs. The capacity of input accumulation as a source of growth is not as limited as that of productivity (at least in the short run), as long as the company does not enter the phase of over-production in which, for example, management may become too stretched and diseconomies of scale may arise.

Ultimately, a rational firm owner would choose the strategy plan of business growth that maximizes their bottom line, i.e., profits. As in GA, output price is normalized at 1 to keep the discussion away from the theory of value, then the problem of profit maximization as it relates to the choice of strategy plan for growth reduces to a problem of cost minimization, which involves input quantities and input prices and the level of TFP. In that case, when implementing output growth through inputs accumulations, a rational firm owner would choose a rational path on the production function where all the input quantities remain proportionately constant to other input quantities, given a fixed set of input prices and a fixed level of TFP. This path is known as Output Expansion Path (OEP) of the firm in economics. The two following figures (Figure 5 and Figure 6) illustrate a typical example of an OEP for a firm having a Cobb–Douglas production function with the traditional inputs of $L$ and $K$ in economics, i.e., Labor and Capital.

Given that a fixed set of relative input prices remains constant over time, a rational business operating already with an optimal combination usage of inputs (aka optimal technology) will continue to choose proportionately the same mixture ratio of inputs when growing output but at a higher and higher level of inputs each time production is to increase to a higher level of output. Moving up along the OEP, output ($Y$) will increase proportionately to the growth in all the inputs equally, as long as there is not a significant change in the real (relative) input prices. In case of a change in the relative input prices, when there is a possibility of technical substitutability between the inputs (for example, if labor can be substituted by capital or vice versa), then some changes and deviations from the above straight OEP will occur towards a more usage of the substitutable input that is then relatively less expensive to mitigate the input price increases, which is sometimes referred to as a change in the technology of production.

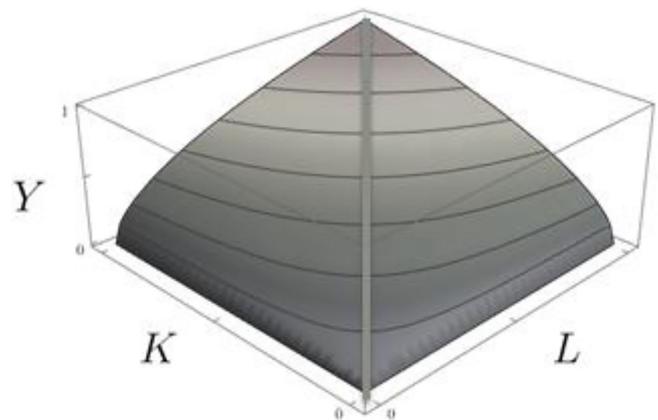

**FIGURE 5** | Output expansion path as a straight line on a three-dimensional Cobb–Douglas production function with two inputs of labor and capital included in the production function. *Source:* Author's Own Drawing.

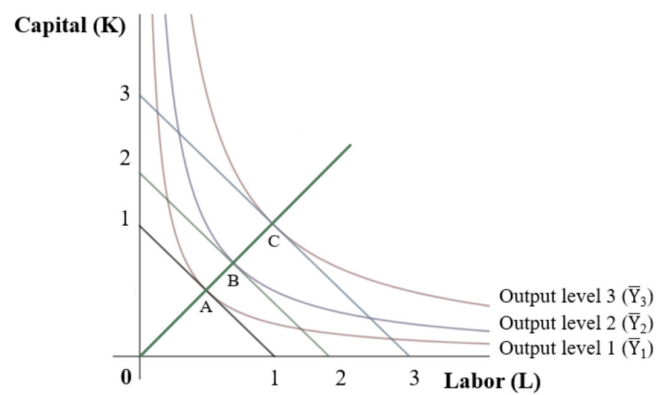

**FIGURE 6** | Output expansion path as a straight line on a two-dimensional space with two inputs of labor and capital included in the production function. *Source:* Author's Own Drawing.

- **Step 7:** According to the parameters of the production function estimated, the set target growth rate, and the growth strategy determined in the previous steps, the required rates of growth in input and output in each year (to achieve the overall target growth over the set timeframe) are computed in this step using the fundamental equation of growth accounting. By doing so, one is working backward through the GA procedure to find the input levels and output level needed in each year in order to ensure the goal achievement within the set timeframe.[8]

- **Step 8:** Finally, in this step, the business growth plan formulated above is implemented, and its performance in achieving the objectives is evaluated periodically, the realized growth rates are compared with the target output growth set for the end of each year, inputs' growths are tuned up and the pace of their increases are adjusted or reconsidered if needed to improve the performance of the output growth plan.

Appendix 2 provides a real-world example of planning the supply side of a business growth using the BGA approach for a propane-delivering company to illustrate how the BGA approach can be put into practice in planning business growth.

The bottom line is that it will be helpful to use the BGA approach introduced in this paper to navigate the process of business growth systematically (on the basis of economic logic and mathematical tools). As shown above, when considering increasing a company's output, it is more logical to begin with the end goal in mind and work backwards, instead of proceeding aimlessly forward and hoping for positive results. This backward approach enables the setting of attainable objectives and meaningful milestones for the following one, three, and five-year periods leading up to the ultimate timeframe targeted.

While BGA offers a systematic, structured, and theoretically grounded approach to planning business growth, it is not without limitations. A significant constraint lies in its dependence on the accuracy and reliability of the input data. The results of the BGA framework hinge on precise measurements of input quantities and output levels, as well as accurate estimations of production function parameters. As with any types of models, any inaccuracies in data collection, recording, or interpretation can lead to erroneous conclusions and suboptimal growth plans. For instance, if the data on labor, capital, or other factors of production are not reported consistently or comprehensively, the estimated growth requirements may fail to reflect the true dynamics of the production process. This dependence on accurate data underscores the importance of meticulous data collection and validation processes before applying the BGA methodology.

Another limitation of BGA can be its reliance on the assumptions underlying the production function used in the analysis. For instance, the commonly employed Cobb–Douglas production function assumes the property of constant returns to scale and specific elasticities of substitution between inputs. While these assumptions are often reasonable approximations, they may not accurately reflect the realities of all types of industries or firms. Deviations from these assumptions, such as varying (increasing or decreasing) returns to scale or the presence of input complementarities, could lead to biased or complicated results. Furthermore, the assumption of stable parameter values over time may not hold in highly dynamic environments where technological changes or shifts in market conditions significantly impact input–output relationships. Recognizing these potential discrepancies, practitioners are encouraged to test alternative functional forms and validate assumptions against empirical evidence.

The BGA framework also requires significant data, expertise, and managerial commitment, which make it less accessible for smaller industries and firms with limited data availability. The extensive data requirements and the need for advanced econometric techniques to estimate production functions can pose obstacles for businesses with constrained resources. Additionally, the iterative nature of the BGA procedure, which involves frequent evaluations and adjustments, demands significant managerial commitment and expertise. To mitigate these challenges, firms may consider seeking external expertise to support its implementation. Despite these challenges, BGA remains a valuable tool when its limitations are acknowledged and complemented with other planning methods and/or external expertise, provided that its application is accompanied by careful consideration of its assumptions and constraints.

Several strategic planning methods provide valuable alternative frameworks for analyzing business growth, each with unique strengths and limitations. SWOT Analysis offers a straightforward assessment of internal and external factors but lacks the quantitative rigor of methods like BGA. Balanced Scorecard (BSC) aligns activities with long-term goals through performance metrics but focuses on execution rather than initial growth planning. Scenario Planning explores multiple future scenarios for flexibility but can be speculative and time-intensive. Portfolio tools like the Growth-Share Matrix prioritize resource allocation but ignore productivity and efficiency. OKRs (Objectives and Key Results) drive focus and accountability but lack broader economic modeling. Methods such as Value Chain Analysis and McKinsey's 7S Framework excel in diagnosing internal processes and organizational alignment but often neglect external dynamics and quantitative precision. Competitive analyses like Porter's Five Forces highlight external pressures but lack actionable internal growth strategies. Finally, project management tools such as PERT and CPM are task-specific and do not address comprehensive strategic planning.

Table 1 below compares a set of well-known methods that can be used for the strategic planning of business growth, highlighting their advantages, disadvantages, and how they complement or contrast with BGA's structured, quantitative approach to growth planning.

**TABLE 1** | Comparison of strategic planning methods that can be used for planning business growth: Advantages, disadvantages, and their relationship to BGA.

| # | Strategic planning method | Description | Advantages | Disadvantages | Comparison to BGA |
|---|---|---|---|---|---|
| 1 | Backward growth accounting (BGA) | A data-driven, backward approach using the logic of growth accounting to plan business growth based on input-output relationships. | Provides a structured, mathematical framework; enables data-driven decisions; an objective approach; identifies key growth drivers. | Requires accurate data and econometric expertise; relies on production function assumptions; less intuitive for non-experts. | N/A |
| 2 | SWOT analysis (SWOT) Learned et al. (1965) | Identifies Strengths, Weaknesses, Opportunities, and Threats to inform strategy, evaluate strategic positions, and guide decisions. | Simple, intuitive, and widely applicable; helps identify internal and external factors affecting strategy; provides a broad strategic overview. | Limited depth; may oversimplify complex situations; too broad of a framework; subjective interpretations can vary; lacks quantitative rigor. | BGA offers more quantitative rigor and specificity in planning, while SWOT provides a broader, more qualitative analysis. |
| 3 | Balanced scorecard (BSC) Kaplan and Norton (2005) | Focuses on aligning organizational activities to the vision and strategy using financial and non-financial metrics (including Financial, Customer, Internal Processes, and Learning and Growth). | Encourages balanced decision-making; integrates financial and operational metrics; aligns activities with goals; focuses on performance measurement and strategic alignment. | Complex implementation; may require significant resources; focuses mainly on alignment, not exploration of alternatives; focuses on implementation rather than initial planning. | BGA focuses on planning future growth with economic logic, while BSC is more focused on operationalizing and tracking performance. |
| 4 | Scenario planning (SP) Wack (1985) | Develops multiple plausible future scenarios to anticipate uncertainties and adapt strategies accordingly. | Encourages strategic flexibility and preparedness for uncertainty; fosters adaptability and resilience; encourages long-term thinking. | Time-consuming, speculative and highly dependent on assumptions and scenario quality; not always actionable; not suitable for short-term planning. | SP addresses uncertainties that BGA assumes to be stable, but it lacks the mathematical structure BGA provides for business growth planning. |
| 5 | BCG growth-share matrix (BCG) Henderson (1970) | Evaluates business units or products based on market growth and market share, helping businesses in planning their portfolio of products and prioritizing investments in their product lines by categorizing products into four quadrants (including Stars, Cash Cows, Dogs, and Question Marks). | Helps prioritize resource allocation among business units or products based on the current, static status (market share) and future, dynamic status (market growth) of products. | Focuses on portfolio management with a greater emphasis on the demand side of business rather than holistic growth planning of the supply side; ignores internal factors like productivity and their planning. | BGA is more operational in addressing both input growth and productivity and can be used for planning each single product separately, while the BCG matrix focuses narrowly on market share and resource prioritization and relative growth planning across products. |

(Continues)

**TABLE 1** | (Continued)

| # | Strategic planning method | Description | Advantages | Disadvantages | Comparison to BGA |
|---|---|---|---|---|---|
| 6 | Objectives and key results (OKR) Grove (1983) | A goal-setting framework used by firms to define measurable goals and track their outcomes. | Drives focus, accountability, and alignment across teams. | Highly reliant on the clarity and quality of objectives; can be overly narrow or granular. | OKR provides goal-oriented alignment, but BGA offers a structured economic framework to plan the growth trajectory systematically, which is a prerequisite for any attempts to check the alignment of outcomes with goals. |
| 7 | Value chain analysis (VCA) Porter (1985) | Examines a business's activities to find value-adding opportunities within all the steps a business takes to deliver a product or service. | Identifies areas to improve efficiency and competitive advantage. | Limited to internal processes and does not address external market or macroeconomic factors. | VCA has a good emphasis on the supply side of business and its internal aspects within the supply chain, offering valuable insights into operational efficiency, but it still lacks the mathematical rigor and the economic logic that BGA brings to business growth planning. |
| 8 | McKinsey's 7S framework (M7S) Waterman, Peters, and Phillips (1980) | Evaluates seven interdependent elements of an organization to ensure they are aligned and working effectively (including Strategy, Structure, Systems, Shared Values, Skills, Style, and Staff). | Focuses on organizational alignment and internal coherence; most effective for facilitating organizational change, improving company's performance, and ensuring cross-departmental alignment. | Limited external market focus; lacks quantitative tools for precise growth planning. | While M7S framework assesses organizational elements and their alignments, BGA provides a more quantitative and actionable roadmap for scaling inputs and productivity for growth. |
| 9 | Porter's five forces (PFF) Porter (1979) | Analyzes industry competition and external threats by assessing competitive forces (including Competitive Rivalry, Supplier Power, Buyer Power, Threat of Substitution, and Threat of New Entry) to understand industry dynamics and develop strategic positions. | Highlights external competitive pressures and market positioning; focuses on competitive positioning; highlights external threats and opportunities in the industry. | Does not provide direct guidance for internal planning or operational decisions; focuses solely on competition; neglects internal factors and the supply side of business; static analysis may not capture dynamic changes, which are critical in planning business growth. | BGA focuses on internal resource planning for growth and the supply side of business growth, while PFF primarily addresses external competitive environments and the demand side of business growth. |

(Continues)

**TABLE 1** | (Continued)

| # | Strategic planning method | Description | Advantages | Disadvantages | Comparison to BGA |
|---|---|---|---|---|---|
| 10 | PERT/CPM (project management tools) Malcolm et al. (1959) | Breaks down tasks and activities and timelines for large projects. | Effective for detailed project planning and execution and time management. | Limited to project management, focuses on the alignment of project activities, and does not address broader strategic or economic goals. | PERT/CPM offers task-level precision but lacks the macro-level perspective of BGA in planning long-term business growth. |
| 11 | McKinsey's strategic horizons (MSH) Baghai, Coley, and White (1999) | Divides growth initiatives into three horizons: maintaining the core business, nurturing emerging business opportunities, and creating genuinely new lines of business. | Emphasizes longer-term growth and innovation by creating future revenue streams, avoiding the pitfalls of short-term profit prioritization. | May reduce focus if resources are spread too thin across horizons; subjective judgment required for horizon categorization. | Can complement BGA by emphasizing phased growth strategies over time, while BGA focuses on structured input–output relationships for immediate and planned growth. |
| 12 | Ansoff matrix (AM) Ansoff (1957) | Helps businesses strategize growth by considering Market Penetration, Market Development, Product Development, and Diversification. | Simplifies growth planning by categorizing strategies into clear quadrants; useful for high-level decision-making. | Overly broad and simplistic; lacks operational guidance or focus on internal capabilities like productivity or resource efficiency. | BGA offers deeper quantitative and operational insights about productivity and resource efficiency, while AM is more of a high-level strategic tool that is more suitable for selecting growth directions. |

The comparison of strategic planning methods highlights the unique advantages of BGA as a tool for business growth. Unlike the other frameworks, BGA provides a mathematically rigorous and theoretically grounded approach, focusing on internal drivers of growth such as input accumulation and productivity improvement. While models like the Ansoff Matrix and Growth-Share Matrix emphasize growth strategies, they primarily focus on the demand side, offering limited insights into the supply-side dynamics which are critical for sustained business expansion. In contrast, BGA specializes in navigating the input–output relationship with a clear roadmap grounded in the production function and the concept of optimal expansion path for rationally and optimally navigating the dynamics of input–output relationship throughout the business growth process. This focus enables practitioners to plan growth dynamically and proportionally, making BGA uniquely positioned to address the complexities of business growth with precision and depth.

Moreover, many other methods, such as SWOT or PFF, prioritize primarily static positioning, external factors, or broad organizational strategies, which, while valuable, lack the specificity and operational focus needed for dynamic growth planning. BGA excels in filling this gap by providing a structured, supply-side-centric framework that can complement these broader methods. It not only accounts for the proportionality of input quantities to output levels but also allows for the rational exploration of alternative growth strategies. This makes it a powerful tool for aligning strategic planning with actionable, data-driven objectives. By integrating BGA into many of these broader frameworks, businesses can create a comprehensive strategy that balances internal growth optimization (on the supply side of business) with external growth outlook (on the demand side of business), enabling an objective and rational determination of the optimal growth path.

All in all, BGA is a valuable tool for business leaders to identify the key factors that drive their business growth. It assists in optimizing utilization and planning for expansion by determining how input quantity and productivity changes affect output growth. Through BGA, analysts can detect deficiencies in the output growth process, such as underutilized labor, inadequate capital investment, or poor productivity. With this information, they can identify the most effective solutions to address these issues. By identifying the critical factors behind output growth and resolving potential shortcomings, businesses can optimize their growth strategy and increase their efficiency. Utilizing BGA as a management tool can help business leaders make informed decisions and take proactive steps towards achieving their growth objectives.

The next section summarizes the main points and provides concluding remarks on the usefulness and applicability of the BGA approach in navigating business growth.

## 6 | Conclusion and Summary

Business growth is an important goal for firms and society as a whole, as it can lead to more profits, stronger financial stability, economic growth and development, more job creation, and possibility of further CSR initiatives. Business growth involves the growth of two distinct dimensions of business operation, namely the demand side and the supply side. Current practices for planning the supply side of business growth are often ad-hoc, and in some cases, businesses operate on a whim with regards to the implementation of their business growth. In light of this, this paper argues that planning business growth can be done more reliably and structurally with the use of the economic-mathematical framework of Growth Accounting (GA).

The GA methodology is routinely used to explain the factors that contribute to economic growth. This involves breaking down growth in output into its sources such as the growth in the "quantity of inputs" and the growth in the "productivity of inputs". This paper proposes a backward approach to GA to help businesses plan for their growth optimally and effectively. The paper discusses the factors critical to consider when planning for growth, how inputs can be structured in an input–output model, and how GA can be modified to suit business operations. The paper introduces an approach called Backward Growth Accounting (BGA), which can be applied to any types of business producing any types of goods or services. The paper provides a formal exposition of GA first, explains its methodology, and provides some examples. Next, it discusses the importance of business growth and planning business growth, introduces the BGA approach afterwards, and concludes with some real-world applications of this approach.

The BGA approach introduced in this paper navigates the output growth process within a logical, mathematical, and realistic framework, and uses an eight-step procedure to create a plan for output growth. In short, the eight steps of the BGA procedure are as follows:

(1) Identification of Input and Output Variables,

(2) Data Collection,

(3) Production Function Specification,

(4) Production Function Estimation,

(5) Setting Objectives & Targets,

(6) Setting a Strategy for Sources of Growth,

(7) Computing Required Year-to-Year Growth in Inputs and TFP, and

(8) Implementation, Evaluation, and Tune-up.

It is suggested that the BGA approach (which is to plan the supply side of business growth) should be considered in conjunction with marketing plans (which are to plan the demand side of business growth) and other comprehensive plans of the business of interest such as strategic plans (which can holistically coordinate all aspects of business operation including the two aspects mentioned already) to ensure consistency, coordination, synergy, financial stability, and economic sustainability.

The main contribution of the present paper is to devise a backward approach to GA (called BGA), which sheds new light on the rational planning of the supply side of business growth by providing a logical, mathematical, realistic, and structural framework for the input–output setting of a production process. The BGA procedure introduced in this paper is a methodology that can be used in business to quantify the impact of various factors on output growth.

This paper advocates that having a theoretically well-established and practically concrete business growth plan on the basis of BGA is certainly far preferable to not having any idea of how and in what proportions different inputs and their productivity levels need to be increased in order to attain a targeted output growth. Of course, changes in market structures, conditions, and outcomes can arise unexpectedly over time, which necessitates that business leaders periodically revise their plans and adjust them based on feedback from the marketplace. By using the BGA approach, businesses can plan for growth in a systematic and data-driven manner, and avoid unreliable, arbitrary, and intuition-driven practices.

Finally, it is important to note that the BGA approach can serve as a valuable management tool for planning business growth, enabling informed decisions and proactive steps towards achieving growth objectives. By identifying key drivers of output growth, BGA helps business leaders optimize utilization for effective expansion planning, while also assessing process shortcomings such as poor productivity, inadequate labor utilization, or insufficient capital investment. With this insight, BGA facilitates the identification of viable solutions to these issues and provides a clear path towards achieving business growth targets. Overall, the BGA method empowers business leaders to make informed, optimal, and strategic decisions to set their business on a path of persistent growth.

---

**Conflicts of Interest**

The author declares no conflicts of interest.

**Data Availability Statement**

Data sharing is not applicable to this article as no new data were created or analyzed in this study.

**Endnotes**

[1] The original famous Solow model which was introduced in Solow (1956) – which was the main aspiration and foundation for the model of GA—was somewhat and somehow discussed around the same time in Swan (1956) as well. As such the original Solow model is also known as Solow-Swan model.

[2] In economics, inputs are also called "resources" or "factors of production".

[3] Put differently, *TFP* refers to the portion of output that is not explained by the amounts of inputs used in product. Rather, its level is determined by how efficiently and intensely the inputs are used, which eventually determines the productivity of factors of production.

[4] The Cobb–Douglas production function is commonly used as the specification of production in growth accounting primarily because of its many desired mathematical properties (e.g., being increasing in its arguments, having continuity, and second-order differentiability), its theoretical support (e.g., being a concave function with respect to each input which represent the law of diminishing marginal product with respect to each input under the ceteris-paribus condition, having convex level curves – aka Isoquant curves – which represent the relative substitutability of inputs but at different tradeoff ratios as the relative availability of an input and thereby its relative importance changes in the process of production, and accommodating different types of returns to scale such as constant returns to scale), and its estimation flexibility (as a multiplicative, power function with a high degree of flexibility to fit many real-world observations well) that this intermediate type of production function provides.

[5] *TFP* has also been called the "level of (production) technology", and accordingly, TFPG has also been called "technical improvement" or "technological progress" in the literature of growth accounting.

[6] *TFP* is in fact a measure of the efficiency with which inputs are used in the process of production. Its growth (TFPG) reflects not only technological progress but also other factors contributing to more output produced, such as improvements in management practices and organizational structure, which altogether are called TFPG.

[7] It is also in this step that the Key Performance Indicator (KPI) of the output growth plan is determined. For every goal that is set for a firm, it is crucial to identify key metrics and results that will help in gauging whether the firm is moving in the direction of the set objective. KPIs offer objectives for teams to strive towards, milestones to measure progress in output growth, and valuable information that aids the company in improving decision-making and navigating the growth plan optimally.

[8] By looking at the second rows of the two visual workflows of the GA and BGA procedures, one can simply see that the **GA** procedure **starts with** the collection of data on the levels and growth rates of inputs and output, and **ends with** the *ex-post*, realized shares of contributing factors to output growth. In contrast, the **BGA** procedure **starts with** the *ex-ante*, targeted rate of growth in output and a growth strategy plan to pin down the contributing factors to output growth, and **ends with** the required levels of inputs and TFP in each year to be able to achieve the targeted output growth rate through the set strategy path. In fact, the BGA procedure is a reverse-engineered version of the GA procedure, as it works through the GA procedure backward. As such, this approach to planning the supply side of business growth is called Backward Growth Accounting (BGA) for this reason in this paper.

[9] Needless to say, a proportionate growth in inputs to obtain a certain rate of output growth is only one possible way to achieve the determined output growth. This path of expansion must be followed by a rational firm owner as long as two conditions hold: (1) The combination of inputs (aka the production technology) in the process of production is already set optimally in the current business operation (which technically means there is not already any overutilization or under-utilization of any inputs), and (2) the relative prices of inputs have not changed from the preceding period. If one of these conditions does not hold true, then the business leaders need to re-consider the combination of inputs; however, any new proposed combination of inputs as the optimal combination must still fall on the production function plane, which represents all the technically optimal combinations of inputs for the given output. As such, the production function should be viewed as the roadmap when considering making a change in the combination of inputs in a process of production due to changes in the relative prices of inputs.

[10] The reasons for TFP growth can include a better organization of the existing production factors by reshuffling and reassigning tasks, better scheduling of the customers so that more demand can be fulfilled with the same amount of inputs (for example, through a better smoothing of demand by scheduling excess demand on busy days to less busy weekdays during which your company has some excess

supply capacity), choosing better and more efficient delivery methods and routes for propane delivery trucks, less waste of input propane in the internal supply chain of the company from receiving propane to delivering propane, employing more educated, more skillful, and more productive personnel, employing more efficient and faster truck drivers in delivering and charging on-ground tanks, more efficient usage of land and building spaces, and the like.

[11] It is important to note that because of compounding effect "across TFP and factor accumulations" (which is separate from compounding effect "over time"), when TFP goes up by 5% annually, then the stocks of production factors do not have to go up by 10% annually, and instead, they need to go up only by 9.53% per year to make the whole annual growth rate in output equal to 15%. If one ignores this second compounding effect in the BGA computations, they are subject to slightly overly grow their TFP level or input quantities compared to what is needed to just meet the target output growth.

[12] It is also important to note that the increase in an input quantity does not have to be always proportionate to that in other inputs. Even when moving along a production function which exhibits the property of CRTS, a growth strategy can be framed by assigning different amounts of increase in different inputs. The choice of growth strategy is not unique in many cases. That is to say, the answer to the question of from what sources an output growth target should be planned is not unique in many cases. After all, a rational firm owner would choose the strategy plan of business growth that minimizes costs and thereby maximizing profits.

[13] This discussion is similar in nature to the difference between the 'approximate' Fisher-effect equation and the 'exact' Fisher-effect equation. When one uses the logarithmic transformations of variables to drive an equation by taking derivatives after the transformation (e.g., the mathematical procedure that is used during the derivation of the Fisher-effect equation and also that of the FEOGA, the resulting equation will be an 'approximate' equation, and works best when the extents of the percentage changes are in the proximity of zero, which is usually the case for both the Fisher-effect equation and also that of the FEOGA. To the extent that the percentage changes become far away from zero, the obtained values from these equations become less precise. To see more information about the discussion of differences between the 'approximate' Fisher equation and the 'exact' Fisher equation, you can see Fisher (1907), Fisher (1997), or Cooper and John (2013) for a more contemporary exposition of this concept. Accordingly, the 'exact' form of FEOGA will be as follows: $(1+g_Y) = (1+g_{TFP})(1+\sum_{i=1}^{i=m} g_{X_i})$, where $g_{X_i}$ is the growth rate in input $X_i$ where there are 'm' inputs in the process of production. After all the difference between the results coming from either equation (i.e., the approximate or exact version) is very small and trivial, especially when the rates of change are small, which is very commonly the case for input growth in planning business growth. Thus, one can keep using the approximate version of FEOGA, which seems to be more intuitive to the general audience.

**References**

Abramovitz, M. 1993. "The Search for the Sources of Growth: Areas of Ignorance, Old and New." *Journal of Economic History* 53, no. 2: 217–243.

Ansoff, H. I. 1957. "Strategies for Diversification." *Harvard Business Review* 35, no. 5: 113–124.

Baghai, M., S. Coley, and D. White. 1999. *The Alchemy of Growth: Practical Insights for Building the Enduring Enterprise*. Perseus Publishing.

Bonaccorsi, A. 1993. "What Do We Know About Exporting by Small Italian Manufacturing Firms?" *Journal of International Marketing* 1, no. 3: 49–75.

Çetinkaya, A. Ş., A. Niavand, and M. Rashid. 2019. "Organizational Change and Competitive Advantage: Business Size Matters." *Business & Management Studies: An International Journal* 7, no. 3: 40–67.

Cooper, R., and A. A. John. 2013. *Macroeconomics: Theory Through Applications*. Flatworld Knowledge. 2012 Book Archive.

Crafts N. 2003. *Quantifying the Contribution of Technological Change to Economic Growth in Different Eras: A Review of the Evidence*. Working Paper No. 79/03, Department of Economic History London School of Economics.

Crafts, N., and K. H. O'Rourke. 2014. "Twentieth Century Growth." In *Handbook of Economic Growth*, vol. 2, 263–346. Elsevier.

Crafts, N., and P. Woltjer. 2021. "Growth Accounting in Economic History: Findings, Lessons and new Directions." *Journal of Economic Surveys* 35, no. 3: 670–696.

Field, A. J. 2006. "Technological Change and US Productivity Growth in the Interwar Years." *Journal of Economic History* 66, no. 1: 203–236.

Fisher, I. 1997. *The Rate of Interest*. Edited by W. J. Barber, J. Tobin, R. W. Dimand, and K. Foster. Pickering & Chatto.

Fisher, I., and W. J. Barber. 1907. *The Rate of Interest: Its Nature, Determination and Relation to Economic Phenomena* 354. Macmillan.

Grove, A. S. 1983. *High Output Management*. Random House.

Henderson, B. D. 1970. *The Product Portfolio*. Boston Consulting Group.

Kaplan, R. S., and D. P. Norton. 2005. *The Balanced Scorecard: Measures That Drive Performance*, Vol. 70, 71–79. Boston, MA: Harvard business review.

Lafuente, E., J. C. Leiva, J. Moreno-Gómez, and L. Szerb. 2020. "A Nonparametric Analysis of Competitiveness Efficiency: The Relevance of Firm Size and the Configuration of Competitive Pillars." *BRQ Business Research Quarterly* 23, no. 3: 203–216.

Learned, E. P., C. R. Christensen, K. R. Andrews, and W. D. Guth. 1965. *Business Policy: Text and Cases*. Richard D. Irwin, Inc.

Lee, J. 2009. "Does Size Matter in Firm Performance? Evidence From US Public Firms." *International Journal of the Economics of Business* 16, no. 2: 189–203.

LpGasMagazine.com. 2022. *2022 Top Propane Retailers*. LP Gas. Retrieved on March 25th at https://www.lpgasmagazine.com/2022-top-propane-retailers/.

Malcolm, D. G., J. H. Roseboom, C. E. Clark, and W. Fazar. 1959. "Application of a Technique for Research and Development Program Evaluation." *Operations Research* 7, no. 5: 646–669.

Moen, O. 1999. "The Relationship Between Firm Size, Competitive Advantages and Export Performance Revisited." *International Small Business Journal* 18, no. 1: 53–72.

Pan, W., T. Xie, Z. Wang, and L. Ma. 2022. "Digital Economy: An Innovation Driver for Total Factor Productivity." *Journal of Business Research* 139: 303–311.

Porter, M. E. 1979. "How Competitive Forces Shape Strategy." *Harvard Business Review* 57, no. 2: 137–145.

Porter, M. E. 1985. *Competitive Advantage: Creating and Sustaining Superior Performance*. Free Press.

Solow, R. M. 1956. "A Contribution to the Theory of Economic Growth." *Quarterly Journal of Economics* 70, no. 1: 65–94.

Solow, R. M. 1957. "Technical Change and the Aggregate Production Function." *Review of Economics and Statistics* 39, no. 3: 312–320. https://doi.org/10.2307/1926047.

Swan, T. W. 1956. "Economic Growth and Capital Accumulation." *Economic Record* 32, no. 2: 334–361.

Wack, P. 1985. "Scenarios: Uncharted Waters Ahead." *Harvard Business Review* 63, no. 5: 73–89.

Waterman, R. H., T. J. Peters, and J. R. Phillips. 1980. "Structure Is Not Organization." *Business Horizons* 23, no. 3: 14–26.

## Appendix 1

### A Numerical Example for Growth Accounting (GA)

Consider an economy whose aggregate production function is estimated to be a Cobb–Douglas of the form $Y = L^\alpha \cdot K^\beta$ with a capital elasticity of output estimated to be $\alpha = 1/3$, and a labor elasticity of output estimated to be $\beta = 2/3$ (assuming that $L$ and $K$ are the only two factors of production). Additionally, assume that it has been computed that in this economy aggregate output grows at 5% per year. Over the same period of time (say, one year), it is computed that its capital stock grows by 6% and its labor force grows by 1%. Now, one can use the Fundamental Equation of Growth Accounting (FEOGA) to answer the following questions.

Question: Part a

a. Compute the portion of growth in output, which is due to changes in the stock of factors of production.

Recall the FEOGA, which is as follows:

$$g_Y = g_{TFP} + \alpha \cdot g_L + \beta \cdot g_K$$

The visual below shows the decomposition of different sources of output growth in this equation.

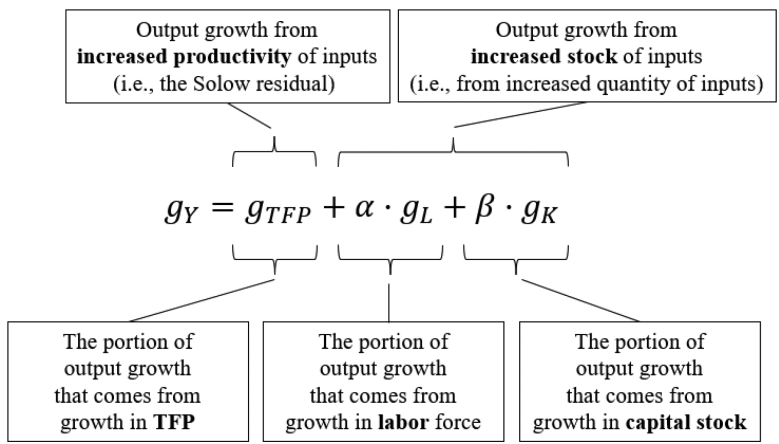

Accordingly, this part of the question is asking to compute and report only the part of the equation that relates to increased stock of inputs, which is: $\alpha \cdot g_L + \beta \cdot g_K$.

$$\alpha \cdot g_L + \beta \cdot g_K = \{0.01 \times (0.67)\} + \{0.06 \times (0.33)\} = 0.0265 \text{ or } 2.65\%$$

This numerical value shows that 2.65% of the 5% economic growth rate in the period under study has occurred just due to the increases in the stock of factors of production ($L$ and $K$).

Question: Part b

b. Compute the portion of growth in output which is due to the increase in the productivity of factors that happened over this period of time (i.e., due to the so-called 'technological progress', or put differently, due to the 'Total Factor Productivity (TFP) growth', which is also known as the Solow Residual).

Now, the FEOGA can be solved for $g_{TFP}$ to answer this question, as shown below:

$$g_{TFP} = g_Y - (\alpha \cdot g_L + \beta \cdot g_K)$$

$$g_{TFP} = 5\% - 2.65\% = 2.35\%$$

This numerical value shows that 2.35% of the 5% economic growth rate in the period under study has occurred just due to the increase in the productivity of factors of production ($L$ and $K$), which is also known as TFPG or the Solow Residual.

(In fact, the results from the preceding part indicated that 2.65% of the 5% economic growth rate during the analyzed period was attributable to the increases in the stock of factors of production ($L$ and $K$). As a result, there remains 2.35% of the growth in output that cannot be explained solely by the increase in the number of workers and the increase in the stock of capital. Indeed, this remaining portion of economic growth has occurred due to the increase in the productivity of factors that has happened over the period of study, or put differently, due to technical progress during this period of time.)

Question: Part c

c. Provide a breakdown of different sources of economic growth in percentage form, attributing different percentage portions of economic growth to different sources of growth.

There are two ways to report the answer to this question: (1) Showing the growth decomposition out of the 5% in terms of 'percentage points', and (2) Showing the growth decomposition out of the 100% in terms of 'percent' (i.e., scaling up the 5% to a scale of 100%). This question is answered below in both ways to avoid possible confusions.

Decomposition of the 5% output growth rate:

$\alpha \cdot g_L = (0.67) \times 0.01 = 0.0067$ or **0.67%** from increased $L$ (i.e., from $g_L$)

$\beta \cdot g_K = (0.33) \times 0.06 = 0.0198$ or **1.98%** from increased $K$ (i.e., from $g_K$)

$g_{TFP} = 5\% - 0.67\% - 1.98\% = \mathbf{2.35\%}$ from increased TFP (i.e., from $g_{TFP}$)

Decomposition in terms of 100%:

$\frac{\alpha \cdot g_L}{g_Y} = \left(\frac{0.0067}{0.05}\right) * 100\% = \mathbf{13.4\%}$ of the total growth has come from increased $L$ (i.e., from $g_L$)

$\frac{\beta \cdot g_K}{g_Y} = \left(\frac{0.0198}{0.05}\right) * 100\% = \mathbf{39.6\%}$ of the total growth has come from increased $K$ (i.e., from $g_K$)

$\frac{g_{TFP}}{g_Y} = \left(\frac{0.0235}{0.05}\right) * 100\% = \mathbf{47\%}$ of the total growth has come from increased TFP (i.e., from $g_{TFP}$)

## Appendix 2

### A Numerical Example for Backward Growth Accounting (BGA) - A Case Study for a Propane-Delivery Company

Real-World Applications of Backward Growth Accounting (BGA) in Planning Business Growth

(Context: A Propane Retailer Competing over National Share with Another Firm)

Propane is a type of gas with a molecular formula of C3H8 that can be compressed into a transportable liquid for use as a fuel in various industrial, residential, and commercial settings. It is a by-product of petroleum refineries and natural gas processing and is considered to be part of a larger group of liquefied petroleum gases (aka LP gases). Propane was initially discovered in France in 1857 and eventually became commercially available in the US in 1911.

According to National Propane Gas Association (NPGA), even though propane users in America might not realize it, they are promoting the use of a clean-burning and low-carbon fuel that plays a critical role in energizing residential, commercial, agricultural, and industrial sectors across all 50 states. The vast majority of the US propane industry comprises small independent firms devoted to catering to the distinct energy requirements of American communities. These businesses frequently operate in regions that have not experienced recent economic development and provide technical training to their staff, which equips them with the skill sets they need, enabling them to be more self-reliant and independent. Because it is a domestically produced alternative fuel, propane creates job opportunities and investment in the American economy, from production to consumption. Over the last decade, production of propane has risen more than twofold. As per the US Department of Energy, propane exports, which stood at no more than 91,000 barrels per day in 2010, averaged nearly 1.1 million barrels per day in 2020, exhibiting a growth of over 1200% in a decade.

According to NPGA, the growth of propane production has created further job openings in different sectors of the propane industry, such as production, transportation, and retail sales and delivery. A 2018 study by the Propane Education & Research Council (PERC) revealed that the propane industry in America employed in excess of 57,000 American workers and generated direct economic impact of $46 billion per year. Furthermore, since these jobs cannot be outsourced, many people find employment security in the propane industry. The economic influence of propane extends to every state and congressional district in America. On a national scale, propane is used in 50 million households across America, with 11.9 million relying on it for space or water heating. The propane industry provides services to 1.1 million commercial accounts, 505,000 agricultural accounts, and 185,000 industrial accounts.

The top three propane retailers in the US are AmeriGas Propane, Ferrellgas, and Suburban Propane, respectively. Table A.1 depicts the annual gallon sales of each of these three companies in 2022, delivered to customers.

In the context of this case study, suppose you are working for Ferrellgas (as an economic-financial planner), which is the second largest firm running behind AmeriGas Propane. Given this context and your position in this company, please answer the following questions.

Question: Part a

a. Assume that, in the past decades, the gallon sales of Ferrellgas that you are working for has grown at the rate of 10% per year. If the gallon sales of the company is expected to keep growing at the same annual rate in the next few decades, **how many years** will it take for your company's gallon sales **to double**? Why? Please show your work. (Hint: You can use the rule of 70 to answer this question.)

$$\# \text{ of yrs for sales to double} = \frac{70}{g_Y(\%)} = \left(\frac{70}{10}\right) = 7 \text{ years.}$$

Question: Part b

b. Assume that, in the past decades, the gallon sales of Ferrellgas that you are working for has grown at the rate of 10% per year. If the gallon sales of the company are expected to keep growing at the same annual rate in the next few decades, **how many years** will it take for your company's gallon sales **to triple**? Why? Please show your work and report the exact answer and not an approximate one. (Hint: You can use the Future Value (FV) formula to answer this question.)

$$FV(Y) = PV(Y)(1+g_Y)^N$$

$$3 * PV(Y) = PV(Y)(1+0.1)^N$$

$$3 = (1+0.1)^N$$

$$Ln(3) = Ln\{(1+0.1)^N\}$$

$$Ln(3) = N * Ln\{(1.1)\}$$

$$N = \frac{Ln(3)}{Ln(1.1)} = \frac{1.099}{0.095} = 11.57 \text{ years}$$

This means it will take almost eleven and a half years for the sales of the company to triple.

Question: Part c

c. The Board of Directors (BOD) of your company (Ferrellgas) has asked the CEO of the company to plan economically and financially in such a way that Ferrellgas becomes the largest company in the propane industry in the US within the next decade and preferably as soon as possible. The CEO has reached out to you as the economic-financial planner of the company and has asked you to provide some if-then scenarios for the growth of your company's gallon sales in the upcoming years, so that the CEO and BOD can consider and select a practical growth path forward given the growth potential in demand for your product and supply capacity of your company.

Recall the discussion about growth in the paper and the discrete equation of growth (i.e., the present/future value formulas). As we can see in Table A1, your major competitor, AmeriGas Propane is running ahead of Ferrellgas in terms of gallon sales (with it being 1097M gallons for AmeriGas vs. 632.057M gallons for Ferrellgas in 2022). According to the historical data and forecast reports that you have collected and created, suppose that it is expected that the AmeriGas' gallon sales will grow annually at the rate of 2% on average and that of Ferrellgas will grow annually at the rate of 9% on average in the next three decades if the current status and pace of your company remains as is now and is considered in the current forecasts. Please use the discrete formula of growth to find out **how many years** it will take for your economy **to overtake AmeriGas** in terms of annual gallon sales. (Hint: To better make sense of this catch-up growth problem, you should draw a real-number line, and put down the four givens that you have above on the real-number line, and use the FV formulas, and solve for N as your unknown).

Find $N$ such that $FV(Ferrellgas) = FV(AmeriGas)$

$$FV_{Fer} = FV_{Ame}$$

**TABLE A1** | The annual gallon sales of the top three national retailers of propane in the US in 2022.

| National propane retailers in the US | | | |
| --- | --- | --- | --- |
| Rank | Company | Location | Gallons |
| 1 | AmeriGas Propane | King of Prussia, PA | 1,097,000,000 |
| 2 | Ferrellgas | Overland Park, KS | 632,057,000 |
| 3 | Suburban Propane Partners | Whippany, NJ | 419,800,000 |

*Source:* https://www.lpgasmagazine.com/2022-top-propane-retailers/.

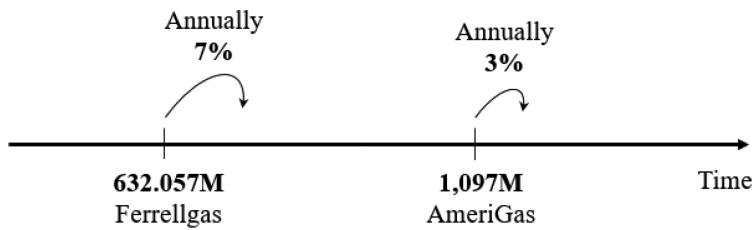

$$PV_{Fer}(1+g_Y^{Fer})^N = PV_{Ame}(1+g_Y^{Ame})^N$$

$$632.057(1+0.07)^N = 1{,}097(1+0.03)^N$$

Using the two well-known properties of the natural logarithm function, which are as follows $Ln(X^N) = N * Ln(X)$ and $Ln(X \cdot Y) = Ln(X) + Ln(Y)$, the above equation can be re-written as follows:

$$Ln(632.057) + Ln(1.07)^N = Ln(1{,}097) + Ln(1.03)^N$$

$$Ln(632.057) + N * Ln(1.07) = Ln(1{,}097) + N * Ln(1.03)$$

$$6.499 + N * 0.068 = 7 + N * 0.03$$

$$N = \frac{0.551}{0.065} = 8.48 \text{ years}$$

This result means that if Ferrellgas's output grows at the annual rate of $g_y^{Fer}$=7% and AmeriGas' output grows at the annual rate of $g_y^{Ame}$=3%, it will take Ferrellgas almost eight years and a half to catch up with AmeriGas in terms of their outputs (i.e., gallons of propane delivered annually).

Question: Part d

d. **At what annual rate** would your company's gallon sales need to increase in order for your company to **achieve its goal** of being the leading company in the industry **only within 5 years** from now?

Find $g_Y^{Fer}$ such that $FV(Ferrellgas) = FV(AmeriGas)$ within $N = 5$ years

$$FV_{Fer} = FV_{Ame}$$

$$PV_{Fer}(1+g_Y^{Fer})^N = PV_{Ame}(1+g_Y^{Ame})^N$$

$$632.057(1+g_Y^{Fer})^5 = 1{,}097(1+0.03)^5$$

There are different ways to solve this equation for $g_Y^{Fer}$. For example, one can use the Newton–Raphson method to find $g_Y^{Fer}$ numerically. Another way to solve for $g_Y^{Fer}$ is to use, for example, the Goal Seek tool of Excel. This equation can also be solved analytically as follows.

Let us call $(1+g_Y^{Fer}) = X$ for simplicity. Then, the above equation can be re-written as the following:

$$632.057 * X^5 = 1{,}097(1.03)^5$$

Using the two well-known properties of the natural logarithm function, which are as follows $Ln(X^N) = N * Ln(X)$ and $Ln(X \cdot Y) = Ln(X) + Ln(Y)$, the above equation can be re-written as follows:

$$Ln(632.057) + 5 * Ln(X) = Ln(1{,}097) + 5 * Ln(1.03)$$

$$6.449 + 5 * Ln(X) = 7 + 0.15$$

$$Ln(X) = \frac{0.701}{5} = 0.1402$$

Using the definition of the natural logarithm function, this equation can be rewritten as follows:

$$X = e^{0.1402} = 1.150$$

$$(1+g_Y^{Fer}) = 1.15$$

$$g_Y^{Fer} = 0.15 = 15\% \text{ per year for 5 years}$$

This result basically means that if Ferrellgas aims to reach their objective of being the largest propane-delivery company in the nation within 5 years from now, they need to grow their output at the annual rate of $g_y^{Fer}$=15% for 5 years, given that AmeriGas' output grows at the annual rate of $g_y^{Ame}$=3%. That way, Ferrellgas will achieve their targeted business growth with the timeframe of 5 years, as set by the company.

**Note:** After completing Part d, for the rest of the questions below, assume that the BOD and CEO have decided to proceed with the **objective of becoming the largest company in the US propane industry in 5 years from now** and want you to plan for this scenario now.

Now, think of your company's operation as an input–output process in which the items listed in the following visual (Figure A1) show the inputs and output of your company's operation.

This input–output setting can be summarized mathematically in a function called production function, as discussed in the paper. The generic form of such a function looks like the following:

$$Y = TFP \cdot f(PI, EM, PT, BD)$$

where the term "TFP" represents a multiplier capturing the level of Total Factor Productivity (TFP) in the operation at hand, while the other right-hand-side variables represent factors of production which are needed to deliver propane to demanders. Furthermore, assume that engineers and economists in your company have estimated the specific form of this production function to be a Cobb–Douglas one with the following form: $Y = TFP \cdot PI^\alpha \cdot EM^\beta \cdot PT^\gamma \cdot BD^\delta$

This production function boils down to the following growth form of the production function as explained in the paper towards the derivation of the fundamental equation of growth accounting:

$$\frac{\Delta Y}{Y} = \frac{\Delta TFP}{TFP} + \alpha \cdot \frac{\Delta PI}{PI} + \beta \cdot \frac{\Delta EM}{EM} + \gamma \cdot \frac{\Delta PT}{PT} + \delta \cdot \frac{\Delta BD}{BD}$$

Which can be re-written as:

$$g_Y = g_{TFP} + \alpha \cdot g_{PI} + \beta \cdot g_{EM} + \gamma \cdot g_{PT} + \delta \cdot g_{BD}$$

where the terms on the right-hand side of this equation are different sources of output growth in this business operation, and each

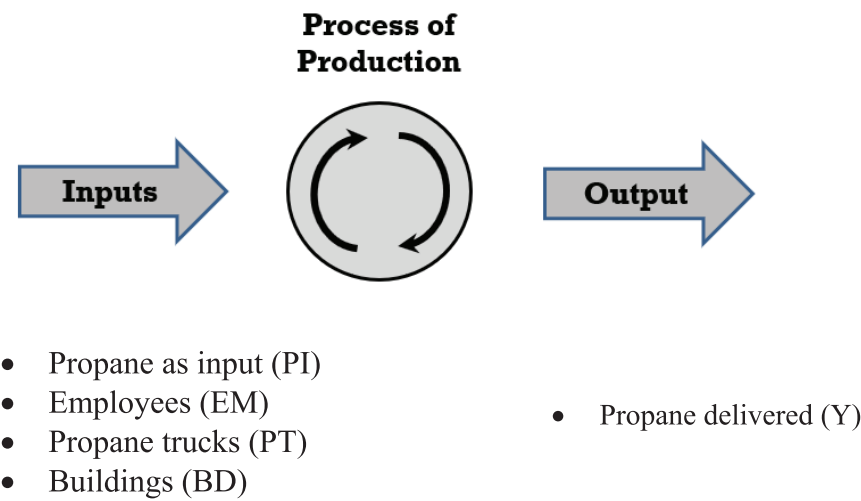

- Propane as input (PI)
- Employees (EM)
- Propane trucks (PT)
- Buildings (BD)

- Propane delivered (Y)

**FIGURE A1** | Input–Output Setting for a Propane Retailer.

parameter represents the relative importance (i.e., the input elasticity of output or factor share) of each respective factor of production, and the term $g_X$ shows the growth in variable X, and when multiplied by its corresponding parameter, the product of the two values indicates the contribution of growth in factor X to growth in output (Y). Additionally, assume that the current technology level parameter (TFP) and factor-share parameters are estimated to be as follows: $TFP = 9811, \alpha = 0.20, \beta = 0.30, \gamma = 0.40, \delta = 0.10$. (Note that these parameters' values add up to 1 (i.e., $\alpha + \beta + \gamma + \delta = 1$), indicating that the underlying production function has the property of Constant Returns To Scale (CRTS), meaning that the company's output (gallons of propane delivered) increases at the same rate as the company's inputs increase, such as labor and capital.

Question: Part e

e. Consider all the information above. Additionally, assume that the BOD and CEO have found it unrealistic to assume that the level of TFP in your company would change in the next 5 years, so the TFP or the productivity level will remain unchanged during the next 5 years (i.e., 0% of the output growth will come from the TFP growth) (that is, $g_{TFP}$ equals 0% during the next 5 years and that all the output growth will come only from factors accumulations), so their strategy with respect to the sources of output growth is to just focus on inputs accumulation. Further, assume that the growth of your business operation occurs optimally on the expansion path of your production function with the structural parameters and relative input prices remaining fixed throughout the growth path. Given all this information, please answer the following question.

 **At what annual rate** do you need to **increase each input** of your business operation to achieve the targeted growth in your output in the next 5 years (in order to end up being the largest company – by gallon sales – in the US propane industry at the end of Year 5), which is the goal set by the BOD and CEO of Ferrellgas? (To answer this question, please construct an Excel table and show the required amount of each input in each year of the next 5 years, so your company can start to plan for those input growths.)

If TFPG is expected to be zero in the next five years, then the only source of output growth that remains is the accumulations of inputs. Considering a proportionate growth in inputs to obtain a certain rate of output growth, that means all inputs need to grow annually at the expected annual rate of growth in output, given that the production process in this business has the property of **CRTS**.[9] Table A2 shows one such possibility where all inputs grow proportionally to other inputs.

The first row of this table provides the givens of the current input–output state of the company at the present time, which includes the current level of production, TFP, propane usage as input (PI), employees (EM) working, propane trucks (PT) being utilized, and buildings (BD) being used. Year 1 (the second row) shows the required levels of inputs and the resulting output in a year from now in such a way that the company can attain its business growth objective within the set timeframe and through the growth strategy set by the BOD and CEO smoothly over five years. As shown in the table, no growth in the productivity of inputs is considered within this 5-year period in this part, and all the output growth will come from a proportionate increase in the quantity of inputs, which is the growth strategy determined by the BOD and CEO in this part of the question.

Question: Part f

f. Given all the information provided before Part e, and now assuming that 5 percentage points of your output growth will come from an increase in the TFP level[10] (i.e., productivity growth) (that is, $g_{TFP}$ constitutes 5 percentage points of $g_Y$) and the remaining amount of output growth will come from other sources of economic growth, which are factor accumulations, and also assuming that the growth of your business operation occurs optimally on the expansion path of your production function with the structural parameters and relative input prices remaining fixed throughout the growth path, please answer the following question.

 **At what annual rate** do you need to **increase each input** of your business operation to achieve the targeted growth in your output in the next 5 years (in order to end up being the largest company – by gallon sales – in the US propane industry at the end of Year 5), which is the goal set by the BOD and CEO of Ferrellgas? (To answer this question, please construct an Excel table and show the required amount of each input in each year of the next 5 years, so your company can start to plan for those input growths.)

Again, the first row of this table provides the givens of the current input–output state of the company at the present time, which includes the current level of production, TFP, propane usage as input (PI), employees (EM) currently working, propane trucks (PT) being utilized, and buildings (BD) being used. Year 1 (the second row) shows the required levels of inputs and the resulting output in a year from now in such a way that the company can attain its business growth objective within the set timeframe and through the growth strategy set by the BOD and CEO smoothly over five years. In this part, the BOD and CEO have decided to proceed with a growth strategy that considers five percentage points of output growth can be obtained through productivity growth and the remaining percentage points are to be obtained from inputs accumulations. As shown in the table, a 5% growth annually in the productivity level of all inputs is considered through this 5-year period, and the remaining part of the output growth will come from proportionate (9.53%) increases in the quantity of inputs, in such a way that they altogether result in a 15% growth per year in output.[11] By completing the above table, we have worked backward through the GA procedure to find the input levels and output level needed in each year in order to ensure the achievement of the target business growth through the designated strategy and within the set timeframe.[12]

**TABLE A2** | The input and output levels required for each year in the next 5 years for Ferrellgas to become the largest propane-delivery company in the US within 5 years from now given the growth strategy set.

| Year | Y | (ΔY/Y) *100% | TFP | (ΔTFP/TFP) *100% | PI (gl) | (ΔPI/PI) *100% | EM (persons) | (ΔEM/EM) *100% | PT (units) | (ΔPT/PT) *100% | BD (sqft) | (ΔBD/BD) *100% |
|---|---|---|---|---|---|---|---|---|---|---|---|---|
| 0 (Now) | 632,057,000 | — | 9811 | — | 695,262,700 | — | 4000 | — | 2400 | — | 1200,000 | — |
| 1 | 726,865,550 | 15% | 9811 | 0% | 799,552,105 | 15.00% | 4600 | 15.00% | 2760 | 15.00% | 1,380,000 | 15.00% |
| 2 | 835,895,383 | 15% | 9811 | 0% | 919,484,921 | 15.00% | 5290 | 15.00% | 3174 | 15.00% | 1,587,000 | 15.00% |
| 3 | 961,279,690 | 15% | 9811 | 0% | 1,057,407,659 | 15.00% | 6084 | 15.00% | 3650 | 15.00% | 1,825,050 | 15.00% |
| 4 | 1,105,471,643 | 15% | 9811 | 0% | 1,216,018,808 | 15.00% | 6996 | 15.00% | 4198 | 15.00% | 2,098,808 | 15.00% |
| 5 | 1,271,292,390 | 15% | 9811 | 0% | 1,398,421,629 | 15.00% | 8045 | 15.00% | 4827 | 15.00% | 2,413,629 | 15.00% |

Question: Part g

g. Given the information and computations that you have done above as well as the numerical values of the parameters of factor of production that were already provided (i.e., $\alpha = 0.20, \beta = 0.30, \gamma = 0.40, \delta = 0.10$), please **provide a breakdown of different sources of annual output growth** in percentage form, attributing different percentage portions of output growth in Part e to different sources of growth (which includes TFPG growth and accumulation of each input).

Decomposition of the 15% output growth rate per year:

$\alpha \cdot g_{PI} = (0.2) \times 0.0953 = 0.01906$ or **1.906%** from increased *PI* (i.e., from $g_{PT}$)

$\beta \cdot g_{EM} = (0.3) \times 0.0953 = 0.02859$ or **2.859%** from increased *EM* (i.e., from $g_{EM}$)

$\gamma \cdot g_{PT} = (0.4) \times 0.0953 = 0.03812$ or **3.812%** from increased *PT* (i.e., from $g_{PT}$)

$\delta \cdot g_{BD} = (0.1) \times 0.0953 = 0.00953$ or **0.953%** from increased *BD* (i.e., from $g_{BD}$)

$g_{TFP} = 0.05$                **5**% from increased TFP (i.e., from $g_{TFP}$)

The effective annual output growth rate from 5% growth in TFP and 9.53% growth in inputs will be equal to $\{(1.05*1.0953)-1\}=1.15-1=0.15=15\%$.[13] Therefore, total percentage growth in output per year is equal to 15% as planned by the BOD and CEO as their objective and strategy plan.

Question: Part h

h. **Please list up the eight steps of the BGA procedure**, which have been taken in this case study, and briefly mention what part of the answers and calculations are related to each step.

As discussed earlier, the BGA approach introduced in this paper uses an eight-step procedure to create a plan for output growth. The eight steps of the BGA procedure in this case study are as follows:

1. Identification of Input and Output Variables

- Identifying propane as input (PI), employees (EM), propane trucks (PT), and buildings (BD) all as inputs, and propane delivered (Y) as output.

2. Data Collection

- Collecting data on PI, EM, PT, BD, and Y.

3. Production Function Specification

$$Y = TFP \cdot PI^{\alpha} \cdot EM^{\beta} \cdot PT^{\gamma} \cdot BD^{\delta}$$

4. Production Function Estimation

$$TFP = 9811, \alpha = 0.20, \beta = 0.30, \gamma = 0.40, \delta = 0.10$$

5. Setting Objectives & Targets

- <u>Objective:</u> Becoming the largest company in the propane industry (in terms of gallon sales) in the US within five years from now
- <u>Target growth:</u> 15% growth in output annually

6. Setting a Strategy for Sources of Growth

- <u>Growth Strategy 1:</u> Output growth only through a proportionate increase in the quantities of inputs while leaving TFP unchanged.
- <u>Growth Strategy 2:</u> Output growth through two sources: 5% of output growth from TFPG and the remaining amount of growth from a proportionate increase in the quantities of inputs

**TABLE A3** | The input and output levels required for each year in the next 5 years for Ferrellgas to become the largest propane-delivery company in the US within 5 years from now given the second growth strategy set.

| Year | Y | (ΔY/Y)*100% | TFP | (ΔTFP/TFP)*100% | PI (gl) | (ΔPI/PI)*100% | EM (persons) | (ΔEM/EM)*100% | PT (units) | (ΔPT/PT)*100% | BD (sqft) | (ΔBD/BD)*100% |
|---|---|---|---|---|---|---|---|---|---|---|---|---|
| 0 (Now) | 632,057,000 | — | 9811 | — | 695,262,700 | — | 4000 | — | 2400 | — | 1,200,000 | — |
| 1 | 726,865,550 | 15% | 10,302 | 5% | 761,521,235 | 9.53% | 4381 | 9.53% | 2629 | 9.53% | 1,314,360 | 9.53% |
| 2 | 835,895,383 | 15% | 10,817 | 5% | 834,094,209 | 9.53% | 4799 | 9.53% | 2879 | 9.53% | 1,439,619 | 9.53% |
| 3 | 961,279,690 | 15% | 11,357 | 5% | 913,583,387 | 9.53% | 5256 | 9.53% | 3154 | 9.53% | 1,576,814 | 9.53% |
| 4 | 1,105,471,643 | 15% | 11,925 | 5% | 1,000,647,884 | 9.53% | 5757 | 9.53% | 3454 | 9.53% | 1,727,085 | 9.53% |
| 5 | 1,271,292,390 | 15% | 12,522 | 5% | 1,096,009,627 | 9.53% | 6306 | 9.53% | 3783 | 9.53% | 1,891,676 | 9.53% |

7. Computing Required Year-to-Year Growth in Inputs and TFP

- Input levels, TFP levels, and output levels reported in Table A2
- Input levels, TFP levels, and output levels reported in Table A3

8. Implementation, Evaluation, and Tune-up

- Now, the business growth plan can be implemented, evaluated, and tuned up afterwards by comparing the target growth rates reported in the tables and the realized growth rates in practice after the fact.

**Additional Examples of Applications of BGA in Different Sectors**

This appendix provided an example application of BGA for strategic planning of business growth in the propane-delivery industry. However, BGA can be used effectively in a wide variety of industries and businesses. Four additional examples of its application are outlined below.

1. *Healthcare Sector: Optimizing Resource Allocation in Hospitals*

In the healthcare industry, BGA can be instrumental in strategic planning for hospital growth. Hospitals often face challenges in scaling up services to meet growing patient demand while maintaining efficiency and quality of care. Using BGA, hospitals can quantify the input requirements—such as staffing, equipment, and facilities—needed to achieve desired output growth in terms of patient care services. For instance, BGA could help estimate the additional number of medical professionals or specialized equipment needed to handle, say, a 20% increase in emergency room visits. By providing a structured and data-driven framework, BGA enables hospital administrators to plan growth scenarios, evaluate resource allocation strategies, and ensure sustainable expansion of service capacity to catch up with increased demand for healthcare services.

2. *Renewable Energy Sector: Scaling Up Green Energy Production*

The renewable energy sector is experiencing rapid growth driven by global environmental goals. BGA can support companies in this sector by helping them strategize the scaling of energy production from sources like wind, solar, and hydropower. For example, a wind energy firm could use BGA to calculate the required investments in turbine installations, maintenance infrastructure, and skilled labor to achieve, for instance, a 15% annual increase in energy output. By incorporating factors such as technological improvements in turbine efficiency and increases in the number of turbines and other such infrastructures, BGA can provide a roadmap for scaling production efficiently while optimizing resource utilization and maintaining cost-effectiveness.

3. *Retail and E-Commerce: Expanding Operations and Supply Capacity*

In the highly competitive retail and e-commerce industries, businesses frequently seek to expand their market presence and product offerings. BGA can be applied to plan the growth of supply chains, warehouse networks, and customer service capabilities. For instance, an e-commerce company aiming to double its sales in three years could use BGA to assess the necessary expansions in logistics, inventory levels, and digital infrastructure. The framework would also help identify the proportional increases in technological investments, human resources, and productivity improvements required to achieve this growth, ensuring that the scaling process is balanced and always in line with the increasing trend of the demand side of the market for the products and services supplied by the business.

4. *Manufacturing Sector: Boosting Production Capacity*

Manufacturers often face the challenge of increasing production capacity to meet rising customer demand while avoiding inefficiencies. BGA can be used to optimize growth in production facilities by analyzing input requirements such as raw materials, machinery, labor, and energy. For example, an automotive manufacturer aiming to launch a new vehicle line could use BGA to determine the additional inputs needed to meet production targets without disrupting existing operations. By focusing on the input–output relationship and the expansion path along the production function, BGA provides manufacturers with a clear, data-driven strategy framework for scaling operations while maintaining economic efficiency.